\documentclass[aps,prb,preprint,onecolumn,superscriptaddress]{revtex4-2}

\usepackage{graphicx}  
\usepackage{amsmath}   
\usepackage{amssymb}   
\usepackage[version=4]{mhchem}
\usepackage{bm}
\usepackage{tabularray} 
\usepackage{booktabs}  
\usepackage{textcomp}

\begin{document}
\title{Observation of unconventional ferroelectricity in non-moir\'{e} graphene on hexagonal boron-nitride boundaries and interfaces}

\author{Tianyu Zhang}
 \email{tycho19@hku.hk}
\affiliation{Department of Physics, The University of Hong Kong, Pokfulam Road, Hong Kong, China}
\affiliation{HK Institute of Quantum Science \& Technology, The University of Hong Kong, Pokfulam Road, Hong Kong, China}
\author{Yueyang Wang}
\affiliation{Department of Physics, The University of Hong Kong, Pokfulam Road, Hong Kong, China}
\affiliation{HK Institute of Quantum Science \& Technology, The University of Hong Kong, Pokfulam Road, Hong Kong, China}
\author{Hongxia Xue}
\affiliation{Department of Physics, The University of Hong Kong, Pokfulam Road, Hong Kong, China}
\affiliation{HK Institute of Quantum Science \& Technology, The University of Hong Kong, Pokfulam Road, Hong Kong, China}
\author{Kenji Watanabe}
\affiliation{Research Center for Electronic and Optical Materials, National Institute for Materials Science, 1-1 Namiki, Tsukuba 305-0044, Japan}
\author{Takashi Taniguchi}
\affiliation{Research Center for Materials Nanoarchitectonics, National Institute for Materials Science, 1-1 Namiki, Tsukuba 305-0044, Japan}
\author{Dong-Keun Ki}
 \email{dkki@hku.hk}
\affiliation{Department of Physics, The University of Hong Kong, Pokfulam Road, Hong Kong, China}
\affiliation{HK Institute of Quantum Science \& Technology, The University of Hong Kong, Pokfulam Road, Hong Kong, China}

\begin{abstract}
Interfacial interactions in two parallel-stacked hexagonal boron-nitride (hBN) layers facilitate sliding ferroelectricity, enabling novel device functionalities. Additionally, when Bernal or twisted bilayer graphene is aligned with an hBN layer, unconventional ferroelectric behavior was observed, though its precise origin remains unclear. Here, we propose an alternative approach to engineering such an unconventional ferroelectricity in graphene-hBN van der Waals (vdW) heterostructures by creating specific types of hBN boundaries and interfaces. We found that the unconventional ferroelectricity can occur--without the alignments at graphene-hBN or hBN-hBN interfaces--when there are hBN edges or interfaces with line defects. By systematically analyzing the gate dependence of mobile and localized charges, we identified key characteristics of localized states that underlie the observed unconventional ferroelectricity, informing future studies. These findings highlight the complexity of the interfacial interactions in graphene/hBN systems, and demonstrate the potential for defect engineering in vdW heterostructures. 
\end{abstract}

\maketitle

\newpage
Recently, an anomalous resistance hysteresis, termed ``unconventional ferroelectricity'', has been observed in dual-gated hBN-encapsulated graphene devices~\cite{zheng2021unconventional,niu2022giant,zheng2023electronic,Yan2023moire,klein2023electrical,chen2024selective,wang2022ferroelectricity,zhang2024electronic,chen2024anomalous,niu2025ferroelectricity,maffione2025twist,Waters2025anomalous,jiang2025theinterplay,lin2025room,singh2025stacking,Tu2025ferroelectric}. 
This phenomenon is characterized by a relatively large electric polarization that exceeds that of sliding ferroelectricity, anomalous screening of one of the gates, and the electronic ratchet effect. 
These features differ significantly from the well-established sliding ferroelectricity found in parallel-stacked hBN layers~\cite{vizner2021interfacial,wang2022interfacial,yasuda2021stacking}, and pave the way for new device applications utilizing their unique features. 

To date, unconventional ferroelectricity has mainly been reported in hBN-encapsulated Bernal~\cite{zheng2021unconventional,niu2022giant,zheng2023electronic,Yan2023moire,maffione2025twist} or twisted~\cite{klein2023electrical,chen2024selective} bilayer graphene, where the asymmetric moiré potential at one of the graphene-hBN interface was attributed to the source of charge localization, thereby leading to the large resistance hysteresis. On the other hand, several recent studies have shown that the moiré potential may not be necessary~\cite{niu2025ferroelectricity,Waters2025anomalous,maffione2025twist,chen2024anomalous}, nor does the number of graphene layers play a decisive role in the hysteretic behavior~\cite{niu2025ferroelectricity,Waters2025anomalous,jiang2025theinterplay,wang2022ferroelectricity,zhang2024electronic,chen2024anomalous,singh2025stacking,lin2025room,maffione2025twist,Tu2025ferroelectric}. These findings indicate that the electrical polarization in graphene-hBN heterostructures is complex and there can be multiple mechanisms that are responsible for the unconventional ferroelectricity in the system, such as moiré-potential induced charge localization~\cite{zheng2021unconventional,klein2023electrical,Yan2023moire,zheng2023electronic,chen2024selective,zhang2024electronic,jiang2025theinterplay,Tu2025ferroelectric}, co-sliding in ABC-like stacking of hBN-graphene-hBN~\cite{lin2025room}, or in tetralayer graphene~\cite{singh2025stacking}, top and bottom hBN alignment~\cite{maffione2025twist}, or particular types of defects in hBN ~\cite{niu2025ferroelectricity,Waters2025anomalous,chen2024anomalous}.

In this study, we take a different approach: engineering defects in hBN by deliberately placing hBN with edges or line defects in graphene-hBN heterostructures, without aligning graphene and hBN in stacking configuration. To investigate the effects of these hBN edges and defects, we incorporated discontinuous few-layer hBN sheets--whose thickness was characterized using the optical contrast technique~\cite{zhang2024accurate}--in close proximity ($<1$~nm) to, or in direct contact with, the graphene layer in some devices (see, e.g., Fig.~\ref{fig:1} and Supplementary Information). By doing so, we explore whether one can design ``defects'' in hBN to control unconventional ferroelectricity~\cite{niu2025ferroelectricity} and gain deeper understanding of the underlying microscopic mechanism of the unconventional ferroelectricity in non-moiré systems. Additionally, to distinguish our system with asymmetric moiré systems more clearly~\cite{zheng2021unconventional,klein2023electrical,Yan2023moire,zheng2023electronic,chen2024selective,zhang2024electronic}, we mainly used monolayer graphene in this study and fabricated reference devices simultaneously that share the same graphene-hBN interfaces without defects.  

Through systematic studies on multiple devices with different geometry (see Supplementary Information), we found that the unconventional ferroelectricity can occur in non-moiré systems when there exist hBN-hBN interface with line defects or natural hBN edges (Fig.~\ref{fig:1}). Detailed (magneto-)resistance measurements further show that the back gate (BG), situated in proximity to the hBN with defects,
affect charge transport in graphene differently from the top gate (TG), which is proximal to the graphene layer (Figs.~2-5). We have also characterized the hysteretic behavior under various experimental conditions beyond those presented in the main text (see Supplementary Information). Although more studies are needed to identify the precise types of defects in hBN that are responsible for the hysteretic behavior, our findings demonstrate that certain types of defects in hBN can induce the ferroelectric behavior~\cite{niu2025ferroelectricity,Waters2025anomalous,chen2024anomalous} and that one can engineer them by design using vdW stacking methods~\cite{Geim2013van,kim2016van,Liu2016van,Novoselov20162D,Wang2023towards}. More broadly, our study shows that defect engineering in vdW heterostructures can be exploited to realize novel functionalities and properties.

\section{Results}

\subsection{Dual-gate hysteresis}

\begin{figure*}[ht] 
    \centering
    \includegraphics[scale=0.95]{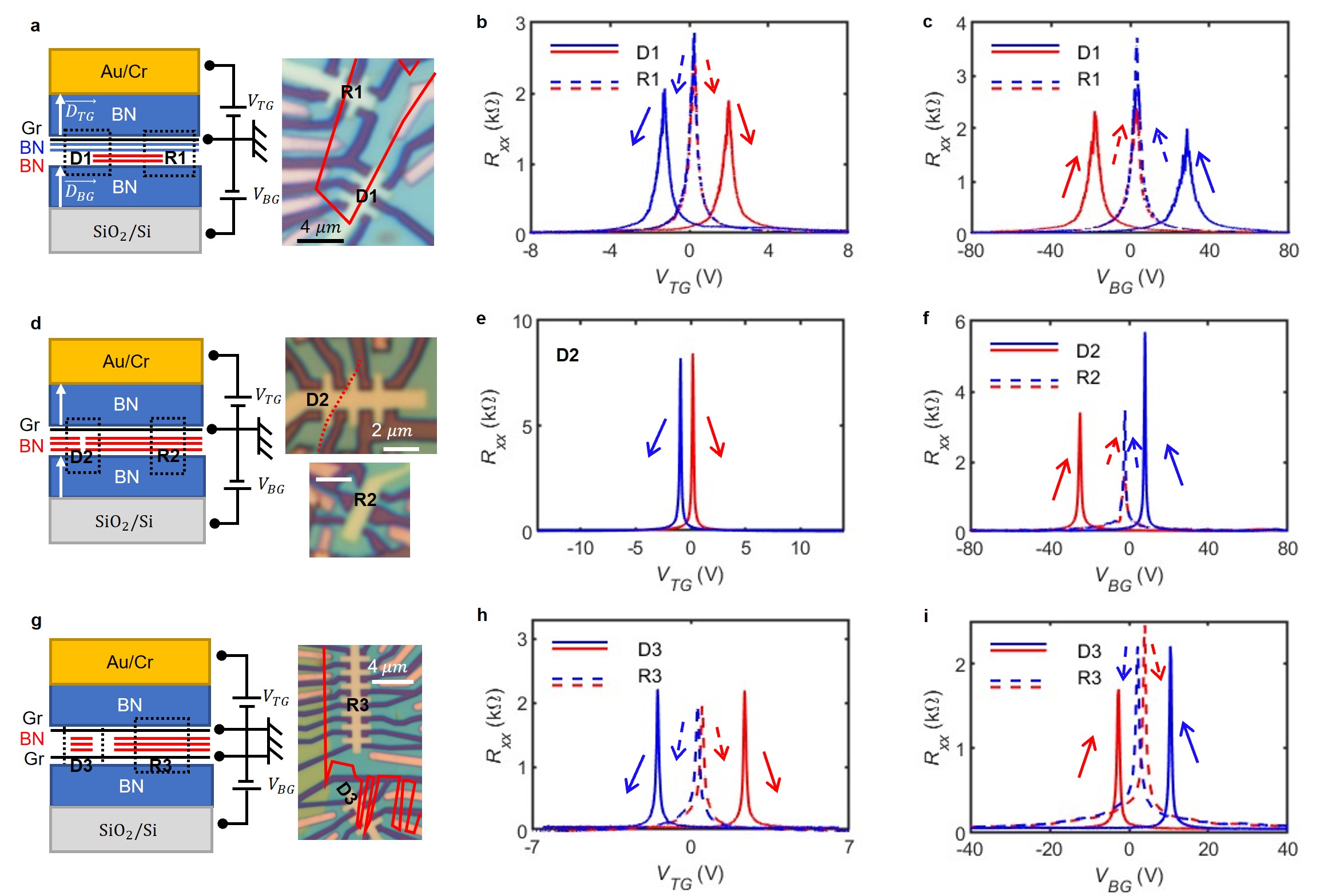}
    \caption{\textbf{Resistance hysteresis in three hBN-encapsulated monolayer graphene devices.} 
    \textbf{a, d, g} Left: Schematics comparing hysteretic devices (D1-D3) with non-hysteretic reference devices (R1-R3). Right: Optical images highlighting different types of hBN defects: a bilayer hBN edges (D1), a trilayer hBN crack (D2), and a trilayer hBN edges (D3). Red lines in the optical images indicate the hBN edges or fracture lines, which are represented by the red-colored lines in the schematic diagrams on the left. \textbf{b, e, h} Resistance $R_{xx}$ as a function of top gate voltage $V_{\text{TG}}$ swept in the forward (red) and backward (blue) directions for devices D1-D3 (solid lines) and R1, R3 (broken lines) at a fixed back gate voltage $V_{\text{BG}}=0$~V. \textbf{c, f, i} $R_{xx}$ in $V_{\text{BG}}$ swept in the forward (red) and backward (blue) directions for devices D1-D3 (solid lines) and R1-R3 (broken lines) at $V_{\text{TG}}=0$~V. Measurements were performed at temperatures of 1.4~K (D1, R1), 43~K (D2, R2), and 5.0~K (D3, R3).}
    \label{fig:1}
\end{figure*}

As shown in Fig.~\ref{fig:1}, we found that some devices with specific types of hBN defects--a crystal boundary (D1), a structural crack (D2), and hBN edges (D3)--exhibit unexpected resistance $R_{xx}$ hysteresis when sweeping both top gate ($V_{\text{TG}}$) and back gate ($V_{\text{BG}}$) voltages in forward and backward directions, characterized by the shift of the resistance peak (i.e., Dirac point) upon changing the sweeping directions (solid lines). On the contrary, the reference samples (R1-R3) that were fabricated at the same time as D1-D3 and share exactly the same graphene-hBN interfaces on both sides of graphene exhibit nearly negligible resistance hysteresis in top and back gate sweeps (broken lines in Fig.~\ref{fig:1}). Notably, except for R1, R2 and R3 have no defects in hBN, signifying the critical role of the hBN defects on the observed hysteresis not the graphene-hBN or hBN-hBN alignments~\cite{niu2025ferroelectricity,Waters2025anomalous}. 

We now compare the results from D1 and R1 shown in Fig.~\ref{fig:1}a as both devices have hBN edges. In the hysteretic device D1, the hBN boundary (red lines) crosses the graphene Hall bar longitudinally while in the reference device R1, the boundary crosses the Hall bar perpendicularly. Even though these two devices share the identical top and bottom hBN layers and the same hBN edge is crossing the Hall bar, only D1 shows a large hysteresis (solid lines in Fig.~\ref{fig:1}b) with the resistance peak shift in density of above $10^{12} \, \text{cm}^{-2}$. In contrast, the reference device R1 exhibits negligible hysteresis. This contrasting result indicates that only specific types of hBN defects or interfaces are responsible for the hysteresis, rather than any types of hBN defects. In the following discussion, we therefore focus on the results from the devices D1 and R1 to characterize the phenomenon.

\subsection{Anomalous screening of the $V_{\text{TG}}$}
\begin{figure*}[ht]
    \centering
    \includegraphics[scale=0.95]{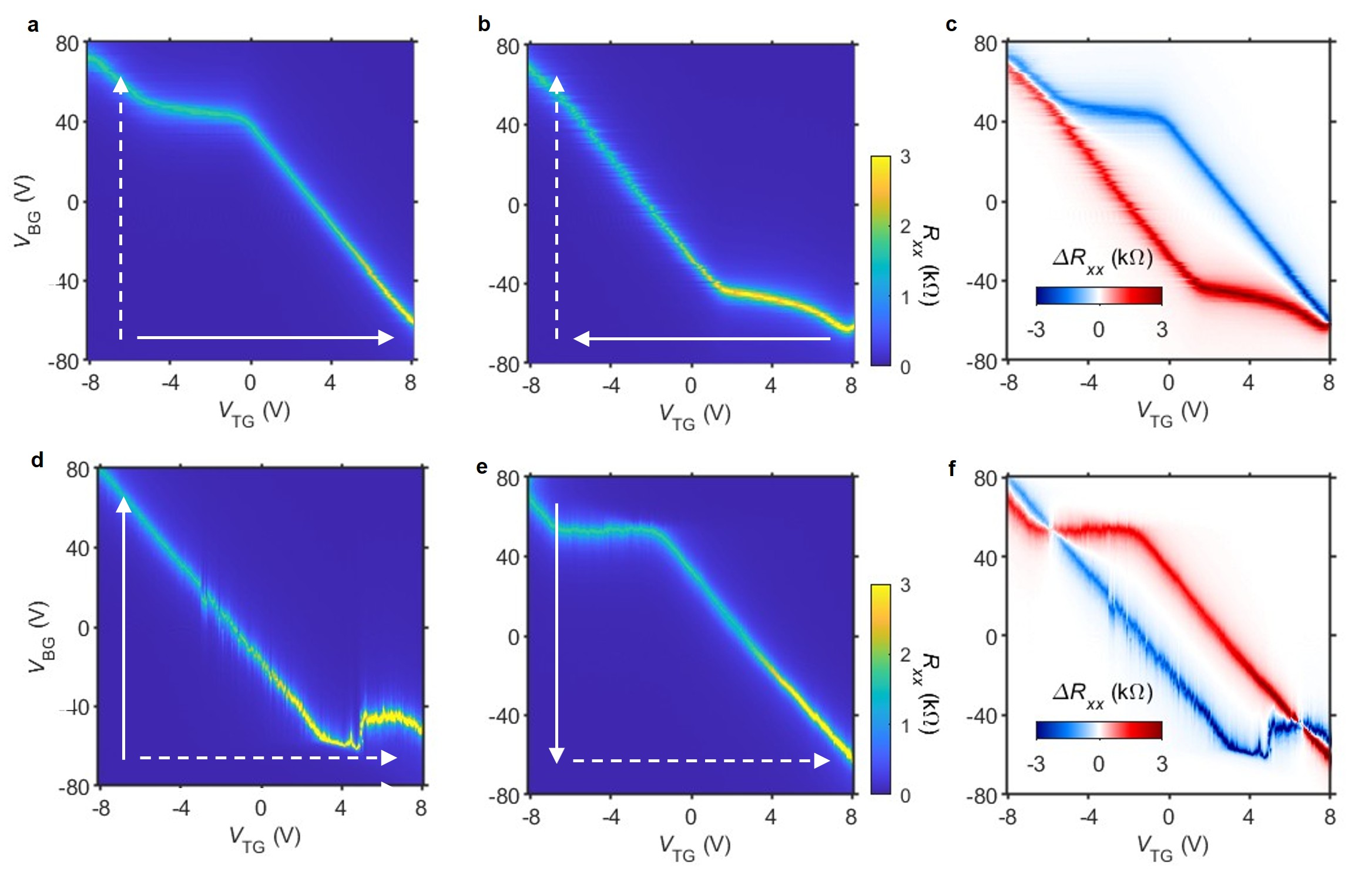}
    \caption{\textbf{Dual-gate resistance hysteresis.} \textbf{a, b} Dual-gate maps of $R_{xx}$ by sweeping $V_{\text{TG}}$ in the forward (\textbf{a}) and backward (\textbf{b}) directions as the fast axis, while gradually ramping up $V_{\text{BG}}$. \textbf{c} The resistance difference between (\textbf{b}) and (\textbf{a}), $\Delta R_{xx}(V_{\text{TG}},V_{\text{BG}})$. \textbf{d, e} Dual-gate maps of $R_{xx}$ by sweeping $V_{\text{BG}}$ in the forward (\textbf{d}) and backward (\textbf{e}) directions as the fast axis, with $V_{\text{TG}}$ slowly increasing. \textbf{f} The resistance difference between (\textbf{e}) and (\textbf{d}), $\Delta R_{xx}(V_{\text{TG}},V_{\text{BG}})$. Solid (dashed) arrows represent the fast (slow) sweep direction.}
    \label{fig:2}
\end{figure*}

To investigate the influence of the top and back gates on the hysteresis in more detail, we measured $R_{xx}$ as a function of both $V_{\text{TG}}$ and $V_{\text{BG}}$. In each measurement, one of the gate voltages was swept forward and backward (``the fast axis'') while the other gate voltage was fixed; the fixed gate was then stepped to a new value after each fast sweep (``the slow axis'').
Figure~\ref{fig:2} shows the results in a color map with the fast (slow) axis marked by solid (dashed) arrows. The figure clearly reveals that the hysteresis is governed by the sweeping trajectories of $V_{\text{TG}}$ and $V_{\text{BG}}$. In contrast to nearly ideal behavior observed in the device R1, where the resistance peak (green-to-yellow in the figure) follows a straight diagonal, zero-density line defined by $n=C_{\text{TG}}V_{\text{TG}}+C_{\text{BG}}V_{\text{BG}}-n_i$ with top and back gate capacitance, $C_{\text{TG}}$ and $C_{\text{BG}}$ respectively ($n_i$: residual impurity density at $V_{\text{TG}}=V_{\text{BG}}=0$~V; see Supplementary Information), the device D1 exhibits a distorted zero-density contour. In particular, we observe the ``horizontal'' features indicating that the modulation by $V_{\text{TG}}$ becomes ineffective. For example, in Fig.~\ref{fig:2}a, the horizontal line appears at $V_{\text{BG}}\approx40$~V during forward $V_{\text{TG}}$ sweeps, whereas it shifts to $V_{\text{BG}}\approx-40$~V during reverse sweeps (Fig.~\ref{fig:2}b). These anomalies imply that a significant fraction of the charges induced by $V_{\text{TG}}$ is immobilized. Similar distorted zero-density lines are observed when $V_{\text{BG}}$ is used as the fast sweep axis (Figs.~\ref{fig:2}d and~\ref{fig:2}e).

In addition, the color maps in Fig.~\ref{fig:2} reveal that during forward sweep of $V_{\text{TG}}$ (Fig.~\ref{fig:2}a), the zero-density line shifts to more positive voltages compared to backward sweeps (Fig.~\ref{fig:2}b). In contrast, when $V_{\text{BG}}$ is used as the fast axis, the shift occurs in the opposite direction, consistent with the opposite hysteresis trace in top and back gate sweeps shown in Fig.~\ref{fig:1}. This opposite behavior is highlighted in Figs.~\ref{fig:2}c and~\ref{fig:2}f where we plot the resistance differences, $\Delta R_{xx}$, between backward and forward sweeps, which becomes positive (red) or negative (blue) when the resistance peak appears in backward or forward sweeps, respectively. These dual-gate hysteresis features differ from sliding ferroelectricity and from many observed unconventional ferroelectricity phenomena~\cite{zheng2021unconventional,niu2022giant,zheng2023electronic,Yan2023moire,klein2023electrical,chen2024selective,niu2025ferroelectricity,maffione2025twist,Waters2025anomalous,jiang2025theinterplay,Tu2025ferroelectric} , with the 
exception of a few cases of dual-gate hysteresis~\cite{wang2022ferroelectricity, 
zhang2024electronic, chen2024anomalous, maffione2025twist, singh2025stacking} 
that do not emphasize the effects of hBN defects.

\begin{figure*}[ht]
    \centering
    \includegraphics[scale=0.95]{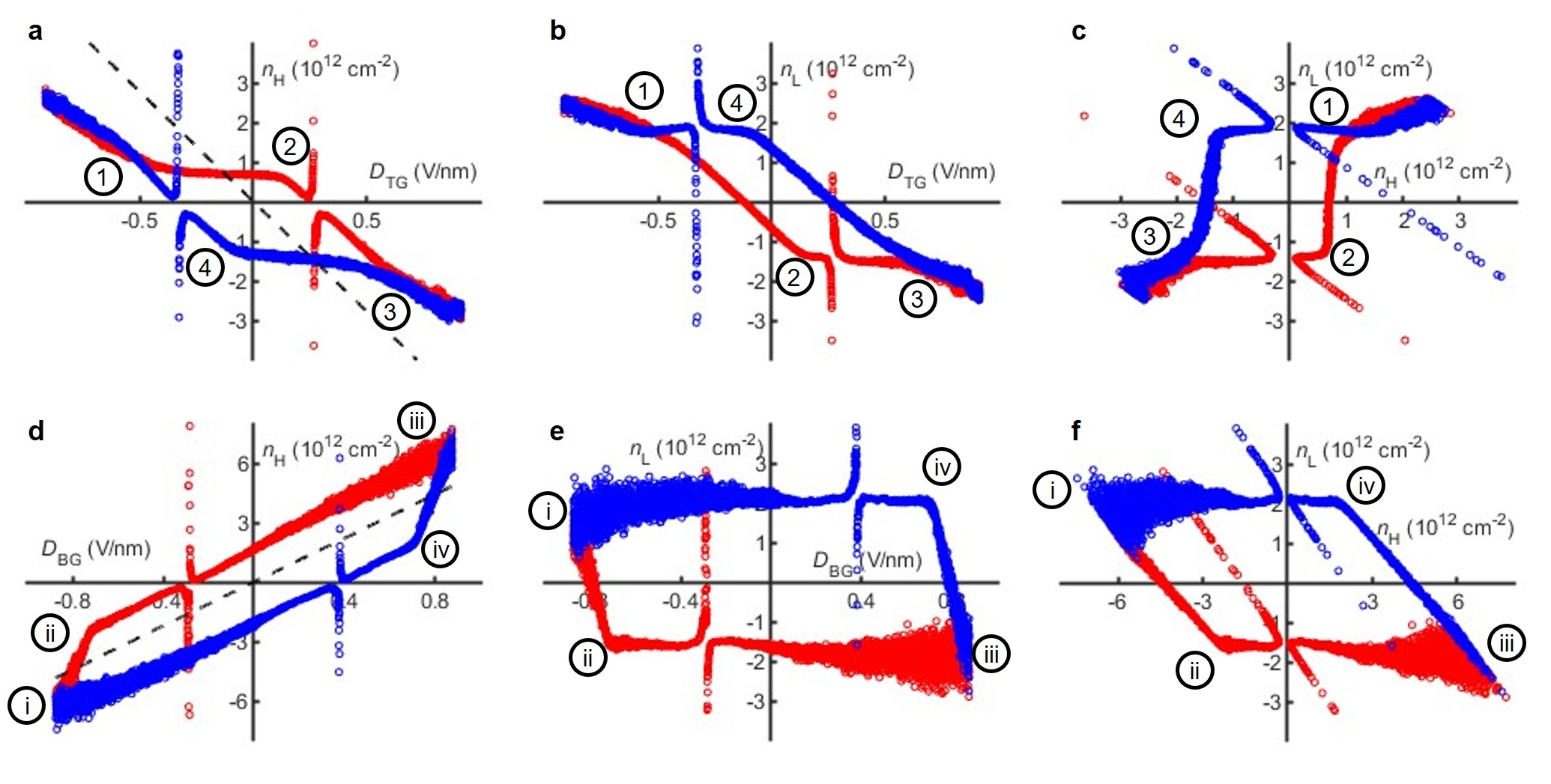}
    \caption{
      \textbf{Density of localized charges $\bm n_{\text{L}}$ and its dependence on $\bm D_{\text{TG}}$ and $\bm D_{\text{BG}}$.} 
      \textbf{a} Hall density $n_{\text{H}}$ versus top gate displacement field $D_{\text{TG}}$  from negative to positive (red), and from positive to negative (blue) at $V_{\text{BG}}=0$~V. The dashed black line represents $n_{\text{tot}} = C_{\text{TG}}V_{\text{TG}}$. \textbf{b} $n_{\text{L}}\equiv n_{\text{tot}}-n_{\text{H}}$ as a function of $D_{\text{TG}}$. \textbf{c}  $n_{\text{L}}$ versus $n_{\text{H}}$ in a sweeping loop of $D_{\text{TG}}$. The sweeping trajectory is indicated in \textbf{a} and \textbf{b}. \textbf{d} $n_{\text{H}}(D_{\text{BG}})$ at $V_{\text{TG}}=0$~V from negative to positive (red), and from positive to negative (blue).  The dashed black line represents $n_{\text{tot}}=C_{\text{BG}}V_{\text{BG}}$. \textbf{e} $n_{\text{L}}(D_{\text{BG}})$. \textbf{f} $n_{\text{L}}$ versus $n_{\text{H}}$ in a sweeping loop of $D_{\text{BG}}$. The sweeping trajectory is indicated in \textbf{d} and \textbf{e}. All Hall measurements are done at $B=0.5\,\text{T}$.}
    \label{fig:3}
\end{figure*}

\subsection{Extraction of localized charges}
Since the anomalous screening of $V_{\text{TG}}$ (Fig.~\ref{fig:2}) indicates a loss of mobile charges, which we denote as localized charges, we quantify their density ($n_{\text{L}}$) to investigate the features in more depth. For this, we first measured the mobile carrier density ($n_{\text{H}}$) through classical Hall measurements in devices D1 and R1. We estimate the gate capacitance, $C_{\text{TG}}=6.27\times10^{11} \,\text{cm}^{-2}\,\text{V}^{-1}$ and $C_{\text{BG}}=5.38\times10^{10}\,\text{cm}^{-2}\,\text{V}^{-1}$, from device R1 that shows negligible hysteresis and shares the same hBN stacks as D1 (see Fig.~\ref{fig:1}a and Supplementary Information). The total carrier density induced by the gates can then be calculated as $n_{\text{tot}}=C_{\text{TG}}V_{\text{TG}}+C_{\text{BG}}V_{\text{BG}}-n_i$ (indicated by the black dashed lines in Figs.~\ref{fig:3}a and \ref{fig:3}d). In our hBN-encapsulated graphene devices (see Supplementary Information), the intrinsic carrier density $n_i$ is approximately $10^{11}\,\text{cm}^{-2}$, which is negligible compared to the gate-induced density and is thus omitted from the calculation in the Device D1. Consequently, the localized charge density can be estimated as $n_{\text{L}} \equiv n_{\text{tot}} - n_{\text{H}}$. Figure~\ref{fig:3} shows the results when sweeping only one of the gates while fixing the other at zero. To discuss the top and back gate responses on equal footing, we express gate voltages in terms of the corresponding displacement fields, $D_\text{TG} = -\frac{C_\text{TG} \cdot V_\text{TG} \cdot e}{\epsilon_0}$ and $D_\text{BG} = \frac{C_\text{BG} \cdot V_\text{BG} \cdot e}{\epsilon_0}$ with elementary charge $e$ and vacuum permittivity $\epsilon_0$ (see Fig.~\ref{fig:1}a for the direction of the displacement field).

Figures.~\ref{fig:3}b and~\ref{fig:3}e show the extracted $n_{\text{L}}$ as a function of $D_{\text{TG}}$ and $D_{\text{BG}}$, respectively, and Figs.~\ref{fig:3}c and~\ref{fig:3}f show the corresponding $n_{\text{L}}$ versus $n_{\text{H}}$ maps. These plots reveal several key features of the hysteretic response of $n_{\text{L}}$ in $D_{\text{TG}}$ and $D_{\text{BG}}$ sweeps. First, the loop is counterclockwise for both sweeps, traced from points~1 to 4 for $D_{\text{TG}}$ (Fig.~\ref{fig:3}b) and from points~i to iv for $D_{\text{BG}}$ sweeps (Fig.~\ref{fig:3}e). Second, at the largest displacement field where the loop begins and ends (corresponding to points~1 and 3 in Figs.~\ref{fig:3}a--c and points~i and iii in Figs.~\ref{fig:3}d--f), the electron carriers ($n_{\text{L}}>0$) are localized at negative $D_{\text{TG}}$ and $D_{\text{BG}}$ while hole carriers ($n_{\text{L}}<0$) are localized at positive displacement fields, independent of the mobile carrier type. Consequently, along $D_{\text{TG}}$ sweep (Fig.~\ref{fig:3}c), $n_{\text{H}}\cdot n_{\text{L}}$ is positive and it is negative along $D_{\text{BG}}$ sweep (Fig.~\ref{fig:3}f). Third, for both sweeps, as the sweep direction reverses, $n_{\text{L}}$ changes its sign before $n_{\text{H}}$ does, and this change persists until $|n_{\text{L}}|$ saturates at $\sim2\times10^{12}\, \text{cm}{^{-2}}$. This behavior is observed between points~1 and 2 and points~3 and 4 in Fig.~\ref{fig:3}c and from points~i to ii and iii to iv in Fig.~\ref{fig:3}f. These observations indicate that the direction of the displacement fields and their sweeping history govern how the localized states are charged, rather than the sign of the gate voltages that set the mobile carrier type.

There are, however, clear differences in top and back gate sweeps as well. In particular, the hysteresis loop for $D_\text{TG}$ sweeps shows a threshold-like onset at $|D_{\text{TG}}|\approx0.6$~V/nm $\equiv D_{\text{TG}}^c$ (corresponding to the points~1 or 3 marked in Figs.~\ref{fig:3}a--c), above which $n_\text{L}$ remains nearly the same for opposite sweep directions. By contrast, the loop made by $D_{\text{BG}}$ sweep (Fig.~\ref{fig:3}d--f) shows no such threshold; $n_{\text{H}}$ and $n_{\text{L}}$ begins to change as soon as the sweep direction is reversed. Because of this difference, the size of the hysteresis loop in $n_{\text{L}}$ versus $n_{\text{H}}$ is larger in $D_{\text{BG}}$ sweep compared to $D_{\text{TG}}$ sweep as for $D_{\text{TG}}$ sweep the hysteresis loop starts only when $|D_{\text{TG}}|<\sim0.6$ V/nm which fixes the $n_{\text{H}}$ at the start of the loop. 
In contrast, during the $D_{\text{BG}}$ sweep, the loop begins as soon as the sweep direction is reversed, allowing $n_{\text{H}}$ to reach a value determined by the maximum $D_{\text{BG}}$, whereas the maximum $|n_\text{L}|$ remains fixed at $\sim 2 \times 10^{12} \, \text{cm}^{-2}$, consistent with the saturation value observed in the $D_{\text{TG}}$ sweep. These findings indicate a different role of the top and back gate displacement fields in modulating the localized states, with the top gate showing a threshold-like onset of the hysteresis while the back gate induces a more instantaneous response for the sweep direction change. 

\subsection{Comparison of top and back gate sweep}
\begin{figure*}[ht]
    \centering
    \includegraphics[scale=1.05]{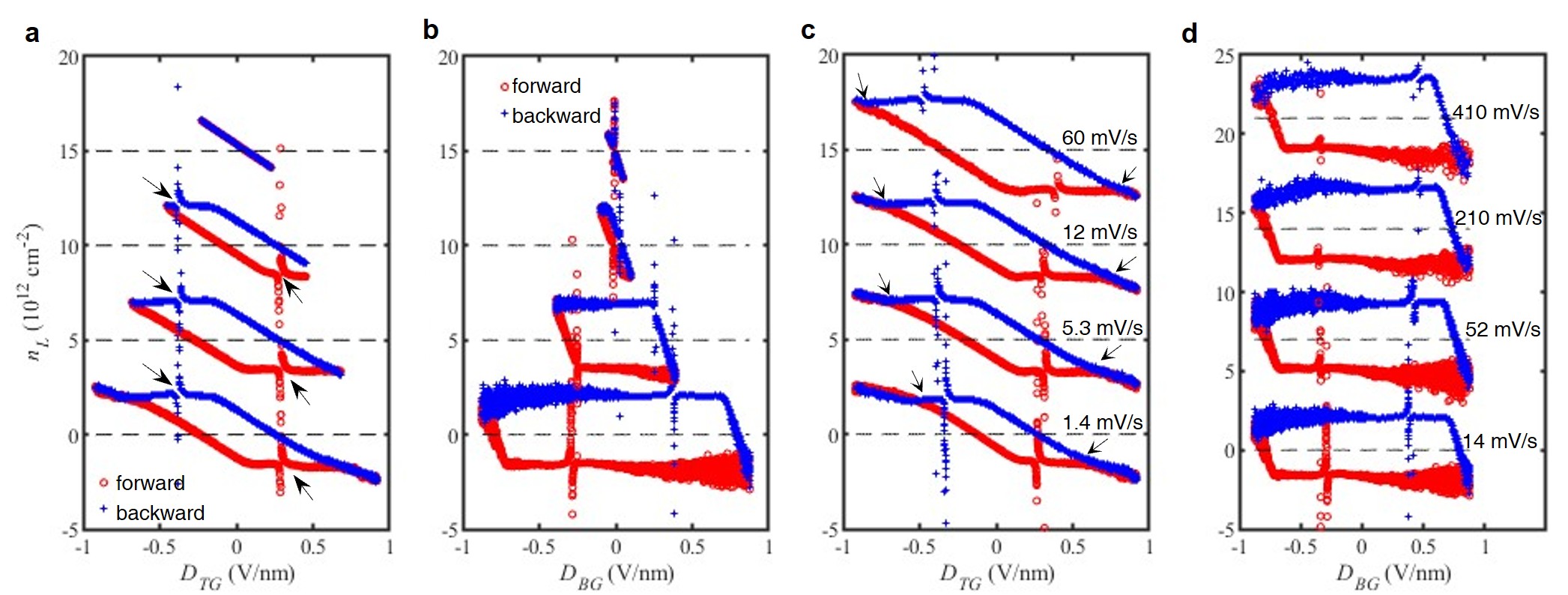}
    \caption{
    \textbf{Dependence of the localized charge density $\bm n_{\text{L}}$ on sweep range and rate.}
    \textbf{a} $n_{\text{L}}$ versus $D_{\text{TG}}$ for forward (negative to positive) and backward (positive to negative) sweeps at different sweep ranges (from bottom to top: $\pm$8.1~V, $\pm$6~V, $\pm$4~V, and $\pm$2~V). The sweep rate is $6.8\,\text{mV/s}$. \textbf{b} $n_{\text{L}}$ versus $D_{\text{BG}}$ for forward and backward sweeps at different sweep ranges (from bottom to top: $\pm$90~V, $\pm$40~V, $\pm$10~V, and $\pm$5~V). The sweep rate is $14\,\text{mV/s}$. \textbf{c} $n_{\text{L}}$ versus $D_{\text{TG}}$ for forward and backward sweeps at different sweep rates. \textbf{d} $n_{\text{L}}$ versus $D_{\text{BG}}$ for forward and backward sweeps at different sweep rates. Curves in \textbf{a--c} are offset by $5\times10^{12} \, \text{cm}^{-2}$ from bottom to top, and curves in \textbf{d} by $7\times10^{12} \, \text{cm}^{-2}$, for clarity.}
    \label{fig:4}
\end{figure*}

\begin{figure*}[ht]
    \centering
    \includegraphics[scale=0.9]{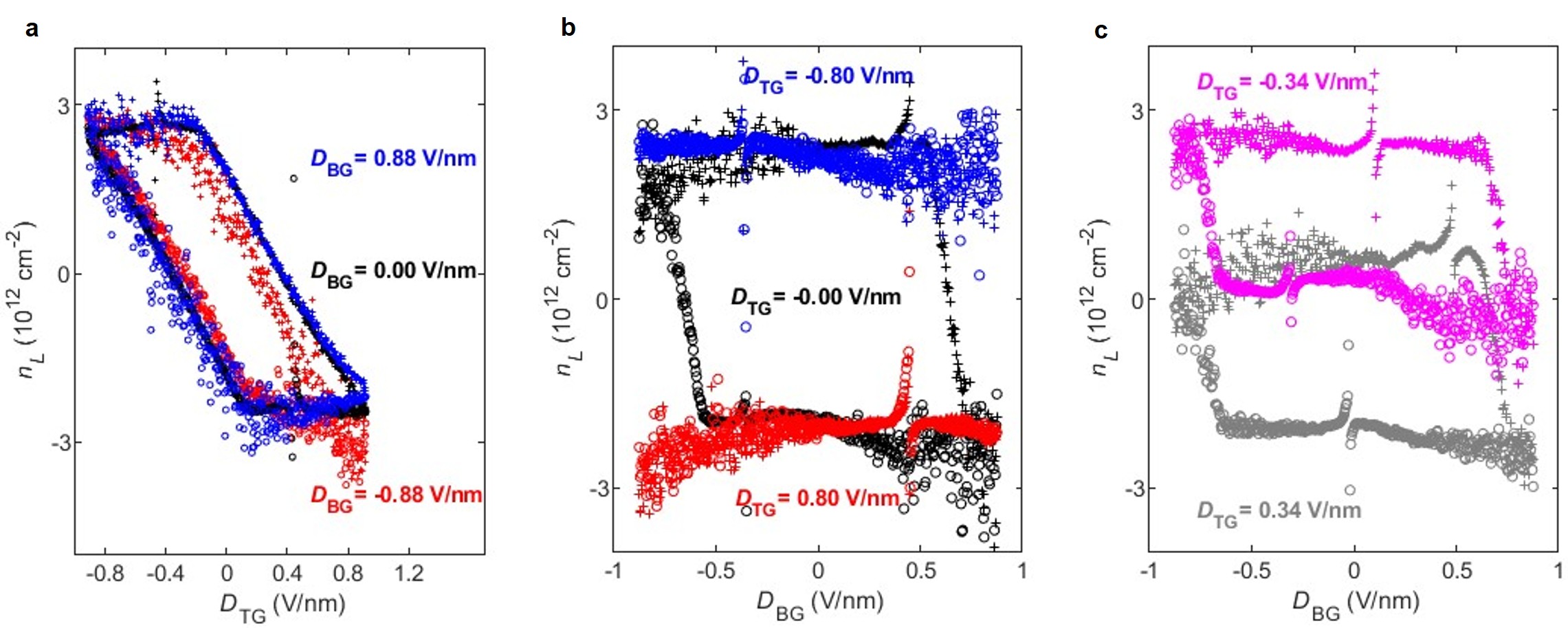}
    \caption{
    \textbf{top and back gate sweep dependent hysteresis.}
    \textbf{a} Localized charge density $n_{\text{L}}$ versus $D_{\text{TG}}$ in forward and backward sweeps at three different values of $D_{\text{BG}}=-0.88$ V/nm (red), 0.00 V/nm (black), and 0.88 V/nm (blue). \textbf{b, c} $n_{\text{L}}(D_{\text{BG}})$ in forward and backward sweeps at five different values of $D_{\text{TG}}=-0.80$ V/nm (blue, \textbf{b}), $-0.34$ V/nm (pink, \textbf{c}), 0.00 V/nm (black, \textbf{b}), 0.34 V/nm (gray, \textbf{c}), and 0.80 V/nm (red, \textbf{b}).}
    \label{fig:5}
\end{figure*}

To investigate the role of the top and back gate sweeps in more detail, we measured the hysteresis loop in $n_{\text{L}}$ as a function of $D_{\text{TG}}$ and $D_{\text{BG}}$ sweeps while varying the sweeping range and the rate. Figures.~\ref{fig:4}a and~\ref{fig:4}b compare the gate-sweep range dependence for top and back gate sweeps, respectively, revealing distinctive differences. When reducing the top gate sweep range (Fig.~\ref{fig:4}a), the Dirac points--characterized by the vertical jumps in the plot marked by black arrows--remain at the same displacement fields. This indicates that the size of the hysteresis loop remains nearly unchanged and reducing the sweep range simply ``cuts" the loop (see Supplementary Information for more details). In contrast, when back gate sweep range decreases (Fig.~\ref{fig:4}b), the Dirac points merges to smaller $D_{\text{BG}}$, indicating that the hysteresis loop gets smaller or even disappears.

Varying the gate-sweep rate (Figs.~\ref{fig:4}c and~\ref{fig:4}d) offers additional insights into the time scale and the dynamics of the localized states charging. For $D_{\text{TG}}$ sweep (Fig.~\ref{fig:4}c), the loop starts at lower $|D_{\text{TG}}|$ as the sweep rate decreases, i.e. the $|D_{\text{TG}}^c|$ marked by arrows decreases, leading to a smaller hysteresis loop. In $D_{\text{BG}}$ sweep (Fig.~\ref{fig:4}d), on the other hand, the loop starts as soon as the sweep direction is reversed independent on the rate. However, the magnitude of the hysteresis window gradually diminishes as the sweep rate is reduced. This behavior indicates that both gates have finite charging and discharging time scales for the localized states, but their dynamics is different when reversing the sweep direction: the back gate responds immediately to change the localized charges while the top gate continues to charge the localized states effectively for a wider range of $|D_{\text{TG}}|>|D_{\text{TG}}^c|$ at a slower sweep rate. It suggests that if we sweep the top gate at a sufficiently slow rate,  hysteresis loop will be removed completely.

In addition, to further probe how top and back gates affect hysteresis, we measured hysteresis loops in $n_{\text{L}}(D_{\text{TG}})$ at different values of $D_{\text{BG}}$ (Fig.~\ref{fig:5}a) and in $n_{\text{L}}(D_{\text{BG}})$ at varying $D_{\text{TG}}$ (Figs.~\ref{fig:5}b and~\ref{fig:5}c). They reveal a distinctive behavior. When sweeping $D_{\text{TG}}$ at fixed $D_{\text{BG}}$ (Fig.~\ref{fig:5}a), the hysteresis loop remains unchanged, in other words, the back gate provides mobile charges only. By contrast, when $D_{\text{TG}}$ is fixed and $D_{\text{BG}}$ is swept (Figs.~\ref{fig:5}b and~\ref{fig:5}c), the top gate directly affects the hysteresis loop and even completely eliminates it at large enough $|D_{\text{TG}}|\approx0.6$ V/nm which is close to $D_{\text{TG}}^c$ at this sweep rate (see Fig.~\ref{fig:4}c and Supplementary Information). This again elaborates distinctive effects of the top and back gate displacement fields on the hysteresis, equivalently on charging and discharging the localized states.

\subsection{Discussions}
From the systematic measurements of $n_{\text{L}}$ in various top and back gate sweep conditions (Figs.~\ref{fig:3}--\ref{fig:5}), we can summarize key characteristics of the hysteresis found in our device:
\begin{enumerate}
    \item The hysteresis loops of $n_{\text{L}}$ in $D_{\text{TG}}$ and $D_{\text{BG}}$ sweeps trace a counterclockwise path, starting from a positive (or negative) $n_{\text{L}}$ at negative (or positive) displacement fields with saturation at $n_{\text{L}}\approx2\sim3\times10^{12}\, \text{cm}{^{-2}}$ (Fig.~\ref{fig:3}).
    \item When the sweep direction is reversed, $D_{\text{BG}}$ changes the localized charges immediately while $|D_{\text{TG}}|$ should become smaller than $D_{\text{TG}}^c$ to start the hysteresis loop (Fig.~\ref{fig:3}).
    \item Saturated $n_{\text{L}}$ and $D_{\text{TG}}^c$ become smaller as the sweep rate decreases (Fig.~\ref{fig:4}).
    \item While $D_{\text{BG}}$ only provide mobile charges when $D_{\text{TG}}$ is swept, $D_{\text{TG}}$ pins partial localized charges when $D_{\text{BG}}$ is swept (Figs.~\ref{fig:5}).
\end{enumerate}
Notably, reference samples possessing identical graphene/hBN interfaces to those of the hysteretic devices did not exhibit the same behavior (Fig.~\ref{fig:1}). Moreover, we fabricated devices with twisted hBN-hBN interfaces without defects (see Supplementary Information) and confirmed that they exhibit a typical sliding ferroelectricity at zero angle (parallel-stacked)~\cite{yasuda2021stacking} or moiré potential effect at a finite angle~\cite{wang2025moire}. These findings strongly suggest that hBN boundaries or defects at hBN--hBN interfaces play a critical role in the observed phenomenon, rather than the moir\'{e} effect~\cite{zheng2021unconventional,zheng2023electronic} or intrinsic defects within the hBN crystal itself~\cite{niu2025ferroelectricity,Waters2025anomalous}. (see Supplementary Information for more discussions).

To identify the possible origin of the localized states in our devices, we have fabricated multiple devices with different types of defects in hBN by design, including i) irregular edges or steps, and ii) hBN-hBN interfaces with the straight edge of the bottom hBN aligned parallel or anti-parallel to the top hBN (see Supplementary Information for a summary). In the devices featuring non-straight hBN edges or steps (i), we found no hysteresis, suggesting that the randomly oriented hBN edges are not the source of the unconventional ferroelectricity. In addition, the device type (ii) with a hBN edge aligned with another hBN also did not exhibit any hysteretic behavior. This is interesting because recently~\cite{fan2025edge}, a small-angle twisted parallel-stacked hBN was shown to host in-plane edge polarization, Then, along the aligned hBN edges, the device type ii hosts potential to create charged domain walls like in three-dimensional ferroelectrics~\cite{sluka2013free}, which may lead to the observed unconventional ferroelectricity. Thus, these two null results imply that the observed hysteresis in our devices may originate from very specific types of hBN edges.

In summary, we presents the results from the systematic studies on non-moire graphene-hBN heterostructures incorporated with hBN boundaries or line defects that reveal strong unconventional ferroelectricity with several unique characteristics, such as different hysteretic behaviors in different gate, a finite time scale in charging and discharging the localized states, and the modulation of hysteresis under the effect of the other gate. By measuring the localized charge density at different gate sweep conditions, we were able to characterize four key properties of the localized states listed above and provide useful insights for future studies. Although the precise mechanism and origin remain unclear, our study demonstrates that hBN edges can play a vital role in inducing unconventional charge polarization in vdW heterostructures. This highlights the importance of careful characterization of hBN flakes and their alignments in vdW stacking to fully understand the underlying mechanism. Additionally and more broadly, our work showcases the potential of vdW stacking as a strategy to engineer defects within structures, enabling the deliberate design and control of novel material properties.

\section{Methods}
\subsection{Device fabrication and measurements}
Graphene and hBN flakes were mechanically exfoliated onto 285~nm $\ce{SiO2}$/Si substrates, while few-layer hBN flakes were exfoliated onto 90~nm $\ce{SiO2}$/Si substrates to facilitate optical identification~\cite{zhang2024accurate}. All devices were fabricated using the conventional dry vdW stacking technique. During the stacking process, there was no deliberate alignments between the graphene and the adjacent hBN layers. The resulting hBN-encapsulated graphene heterostructure was then transferred onto a highly doped $\ce{SiO2}$/Si substrate as the back gate, followed by standard e-beam lithography, metal deposition, 1D edge contacts, and etching to make dual-gated Hall bar devices. Fabricated devices were measured in dry cryogenic systems equipped with superconducting magnets (base temperature: 1.5~K). All measurements were conducted with a 10 nA or 100 nA AC current bias, using a lock-in amplifier at 17.777 Hz and low-noise preamplifiers to minimize noise.

\begin{acknowledgments}
The work is financially supported by the National Key R\&D Program of China (2020YFA0309600) and by the Hong Kong SAR University Grants Committee/Research Grants Council under Area of Excellence schemes (AoE/P-701/20), CRF (C7037-22G) and GRF (17309722 and 17300725). K.W. and T.T. acknowledge support from JSPS KAKENHI (Grant Numbers 19H05790, 20H00354, and 21H05233) and A3 Foresight by JSPS.
\end{acknowledgments}

\section*{Author contributions}
D.K.K. conceived and supervised the project. T.Z. performed sample fabrication, measurements, and data analysis, with assistance from H.X. in fabrication. Y.W. contributed to the fabrication and measurement of device D3. K.W. and T.T. provided the hBN crystals. T.Z. and D.K.K. wrote the paper, and H.X. assisted in revising the manuscript.

\section*{Competing interests}
The authors declare no competing interests.

\section*{Additional information}
\textbf{Correspondence} and requests for materials should be addressed to Tianyu Zhang (email: tycho19@hku.hk) and Dong-Keun Ki (dkki@hku.hk).

\clearpage 
\onecolumngrid 
\begin{center}
\textbf{\large \textnormal{Supplementary Information for}\\ Observation of unconventional ferroelectricity in non-moir\'{e} graphene on hexagonal boron-nitride boundaries and interfaces}
\end{center}
\setcounter{equation}{0}
\setcounter{figure}{0}
\setcounter{table}{0}
\setcounter{page}{1}
\renewcommand{\theequation}{S\arabic{equation}}
\renewcommand{\thefigure}{S\arabic{figure}}
\renewcommand{\thetable}{S\arabic{table}}
\renewcommand{\thepage}{S\arabic{page}} 
\newcommand{\pcenter}[1]{\multicolumn{1}{>{\centering\arraybackslash}p{#1}}}

\makeatletter
\renewcommand{\bibnumfmt}[1]{[S#1]}
\renewcommand{\citenumfont}[1]{S#1}
\makeatother
\Large
\normalsize
\begin{center}
    \begin{enumerate}
    \item[] Supplementary Note 1: Additional data in the device D1
    \item[] Supplementary Note 2: Additional data in the reference device R1
    \item[] Supplementary Note 3: Additional data in the device D2
    \item[] Supplementary Note 4: Additional data in the device D3
    \item[] Supplementary Note 5: Summary of the devices with twisted hBN interface
    \item[] Supplementary Note 6: Summary of the devices with different types of hBN defects
    \item[] Reference
\end{enumerate}
\end{center}

\newpage

\large
\noindent \textbf{Supplementary Note 1: Additional data in the device D1}
\vspace{5mm}

\normalsize
\noindent \textbf{Magnetic transport measurements}

\begin{figure}[ht]
    \centering
    \includegraphics[scale=0.95]{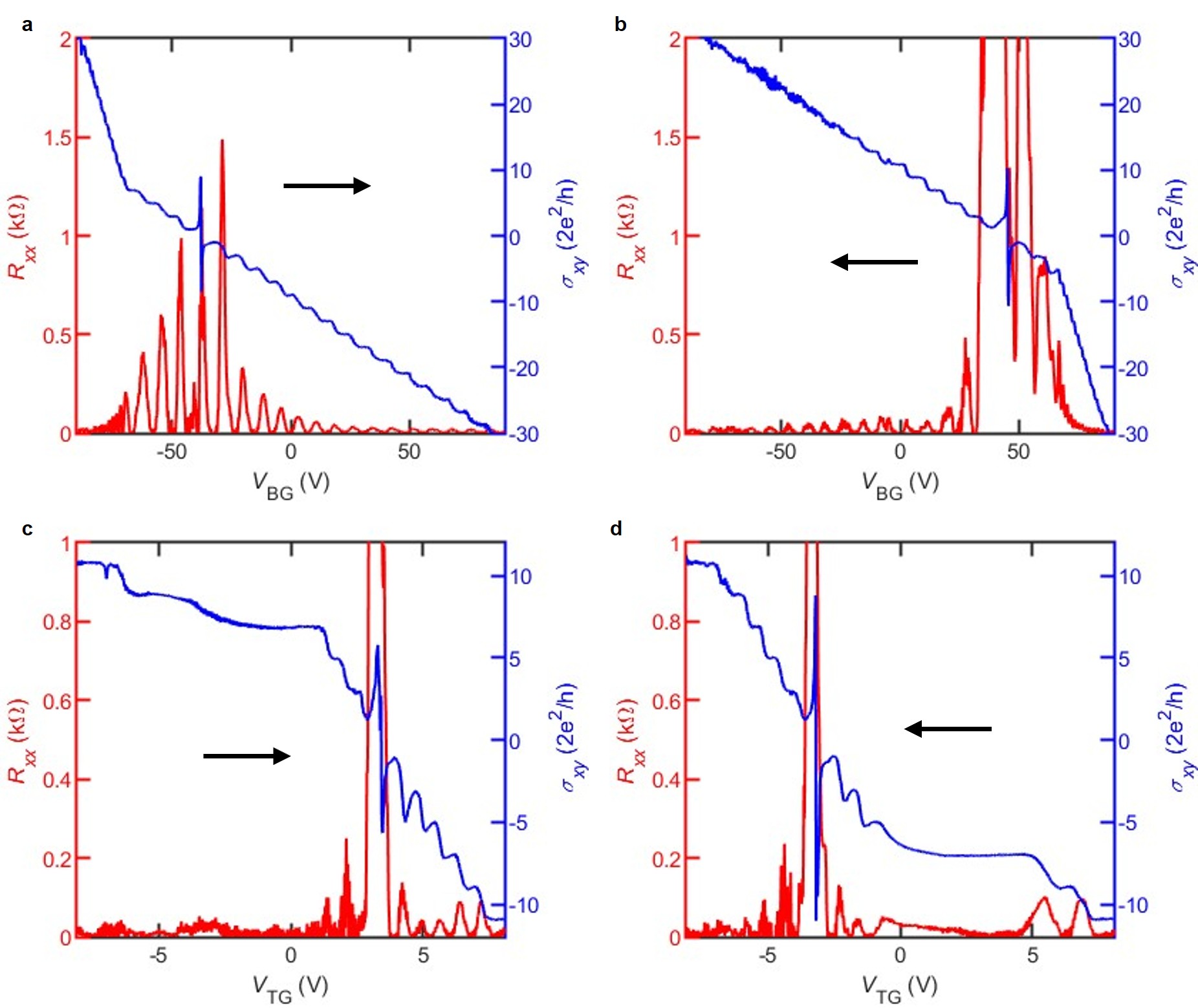}
    \caption{\textbf{Magnetic transport measurement in device D1.} Longitudinal resistance $R_{xx}$ (left axis) and transverse conductivity $\sigma_{xy}$ (right axis) as a function of back gate voltage $V_\text{{BG}}$ in the forward (\textbf{a}) and backward (\textbf{b}) sweep directions. Longitudinal resistance $R_{xx}$ and transverse conductivity $\sigma_{xy}$ (right axis) as a function of top gate voltage $V_\text{{TG}}$ in the forward (\textbf{c}) and backward (\textbf{d}) sweep directions. All measurements are done at B=5~T and T=1.4~K with another gate voltage fixed to 0 V.}
     \label{fig:s1}
\end{figure}

The number of graphene layers was initially identified via optical contrast under an optical microscope~\cite{blake2007making,jung2007simple,ni2007graphene,wang2012thickness} and subsequently corroborated by quantum Hall (QH) measurements~\cite{zhang2005experimental} performed on device D1. As illustrated in Fig.~\ref{fig:s1}, well-defined filling factors of $\nu = \pm 2, \pm 6, \pm 10, \pm 14$, and higher-order plateaus are clearly resolved. These magnetotransport measurements further exhibit hysteretic behavior consistent with the observations presented in Fig.~1b and 1c. Notably, the carrier density remains invariant over specific intervals of the top-gate voltage ($V_{\text{TG}}$) sweep, as evidenced by the elongated stable Hall conductivity plateaus (e.g., $\sigma_{xy} = \pm 14e^2/h$). 

The observation of these elongated $\sigma_{xy}$ plateaus against variations in $V_{\text{TG}}$ suggests that the unconventional ferroelectricity within the system acts as a charge buffer, effectively pinning the carrier density and stabilizing the quantum Hall quantization. This stability is of significant practical interest, as the coupling between ferroelectric polarization and the electronic states of graphene could be exploited to broaden the quantized plateaus. Such a mechanism potentially enhances the robustness of the QH states against external electrical noise or gate fluctuations, which is particularly advantageous for precision metrology and the development of next-generation quantum resistance standards.

Furthermore, the high precision of the $\sigma_{xy}$ plateaus, coupled with the near-vanishing longitudinal resistance ($R_{xx}$), underscores the exceptional electronic quality of the graphene. These results effectively exclude intrinsic graphene defects as the origin of the observed hysteresis. Moreover, the quantitative agreement between the magnetotransport data and the classical Hall measurements confirms the reliability of extracting the mobile carrier density in graphene through standard Hall characterization.
\\

\noindent \textbf{Temperature dependence of the unconventional ferroelectricity}
\begin{figure}[ht]
    \centering
    \includegraphics[scale=1]{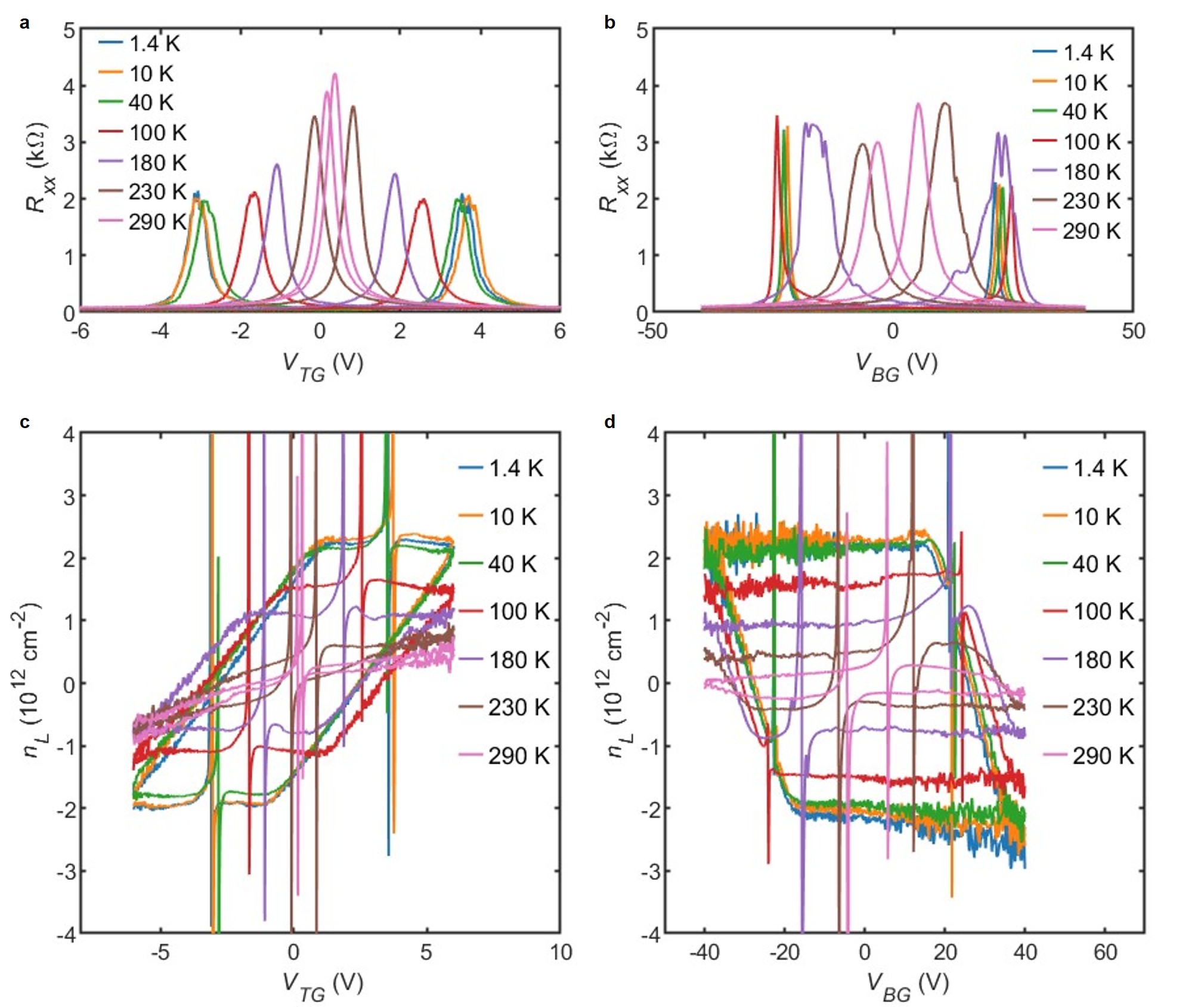}
    \caption{\textbf{Temperature dependence of the unconventional ferroelectricity in device D1.} Longitudinal resistance $R_{xx}$ (left axis) as a function of top gate voltage $V_\text{{TG}}$ (\textbf{a}) and back gate voltage $V_\text{{BG}}$ (\textbf{b}) in the forward and backward sweep direction at different temperature. Extracted localized carrier density $n_{\text{L}}$ as a function $V_\text{{TG}}$ (\textbf{c}) and $V_\text{{BG}}$ (\textbf{d}). All measurements were conducted at magnetic field B=0.5~T with another gate voltage fixed to 0 V.}
     \label{fig:s2}
\end{figure}

As shown in Figs.~\ref{fig:s2}a and \ref{fig:s2}b, the Dirac point shift in $R_{xx}$ between forward and backward sweeps---for both $V_{\text{TG}}$ and $V_{\text{BG}}$---gradually decreases as the temperature is increased from 1.4~K to 290~K. This behavior stands in contrast to the sliding ferroelectricity observed in parallel-stacked hBN, where the hysteresis amplitude (defined as the Dirac point shift between forward and backward gate sweeps) is temperature-independent. In such systems, the electric polarization is intrinsically determined by the hBN stacking configuration rather than thermally activated processes~\cite{yasuda2021stacking_s}. 

Furthermore, as illustrated in Figs.~\ref{fig:s2}c and \ref{fig:s2}d, the hysteresis window of the localized charge density ($n_{\text{L}}$) remains nearly constant up to 100~K. This suggests that the charge localization remains stable under thermal fluctuations up to approximately 40--100~K, which corresponds to an estimated potential depth for the localized states in the range of 3.4 to 8.6~meV. At temperatures approaching room temperature, both the hysteresis window and the magnitude of $n_{\text{L}}$ are significantly reduced. This observation provides direct evidence that localized charges are thermally activated from these potential wells into the graphene Dirac cone, where they subsequently behave as mobile carriers.

\vspace{10mm}
\noindent \textbf{Range dependence of Hall density}
\begin{figure}[ht]
    \centering
    \includegraphics[scale=1]{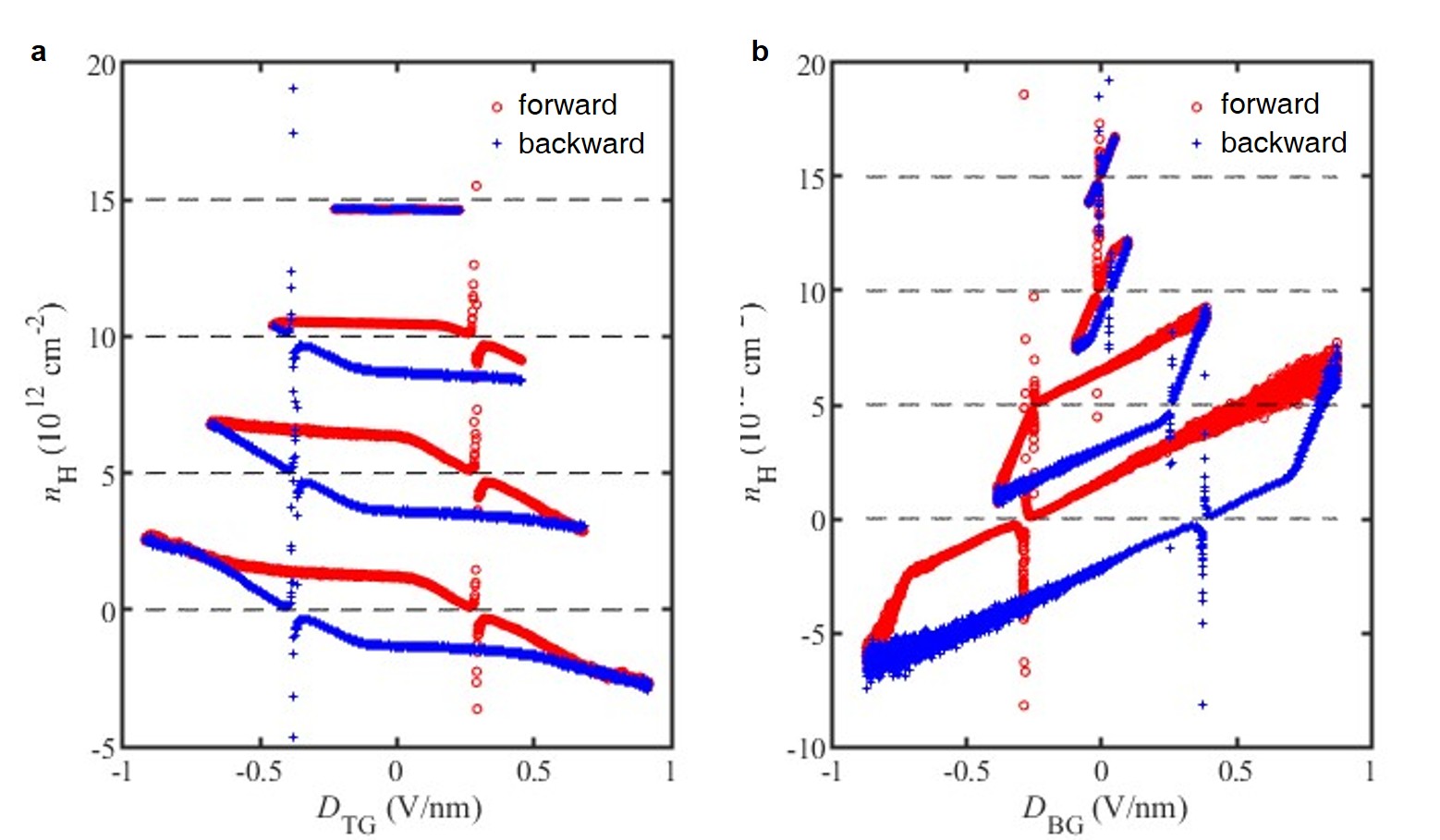}
    \caption{\textbf{Hall density over different sweeping range of top- and back-gate in device D1.} $n_{\text{H}}$ versus $D_{\text{TG}}$ for forward and backward sweeps at different sweeping ranges (from bottom to top: $\pm$8.1~V, $\pm$6~V, $\pm$4~V, and $\pm$2~V). The sweeping rate is $6.8\times10^{-3}\,\text{V/s}$. \textbf{b} $n_{\text{H}}$ versus $D_{\text{BG}}$ for forward and backward sweeps at different sweeping ranges (from bottom to top: $\pm$90~V, $\pm$40~V, $\pm$10~V, and $\pm$5~V). The sweeping rate is $1.4\times10^{-2}\,\text{V/s}$. All measurements were conducted at magnetic field B=0.5~T with another gate voltage fixed to 0 V.}
     \label{fig:s3}
\end{figure}

As shown in Fig.~4a and Fig.~\ref{fig:s3}a, when the sweep range of $V_{\text{TG}}$ decreases from 8.1~V (bottom curve) to 4~V (second-from-top curve), the hysteresis amplitude in $n_{\text{L}}$ versus $D_{\text{TG}}$ remains nearly unchanged, as characterized by the Dirac point shift between forward and backward sweeps. This indicates that the maximum stored localized charge are nearly identical across these sweep ranges of $D_{\text{TG}}$, consistent with the same size of hysteresis window of $\sim 3.7\times10^{12}\,\text{cm}^{-2}$ in the three curves (extracted from the difference between the stable maximal and minimal $n_{\text{L}}$). However, when the sweep range of $D_{\text{TG}}$ is too small---such as a $V_{\text{TG}}$ range of 2~V (top curve in Fig.~\ref{fig:s3}a)---all electrostatically doped charges enter the localized states, as evidenced by the unchanged Hall density $n_{\text{H}}$ under $D_{\text{TG}}$ sweep range.

\begin{figure}[ht]
    \centering
    \includegraphics[scale=1]{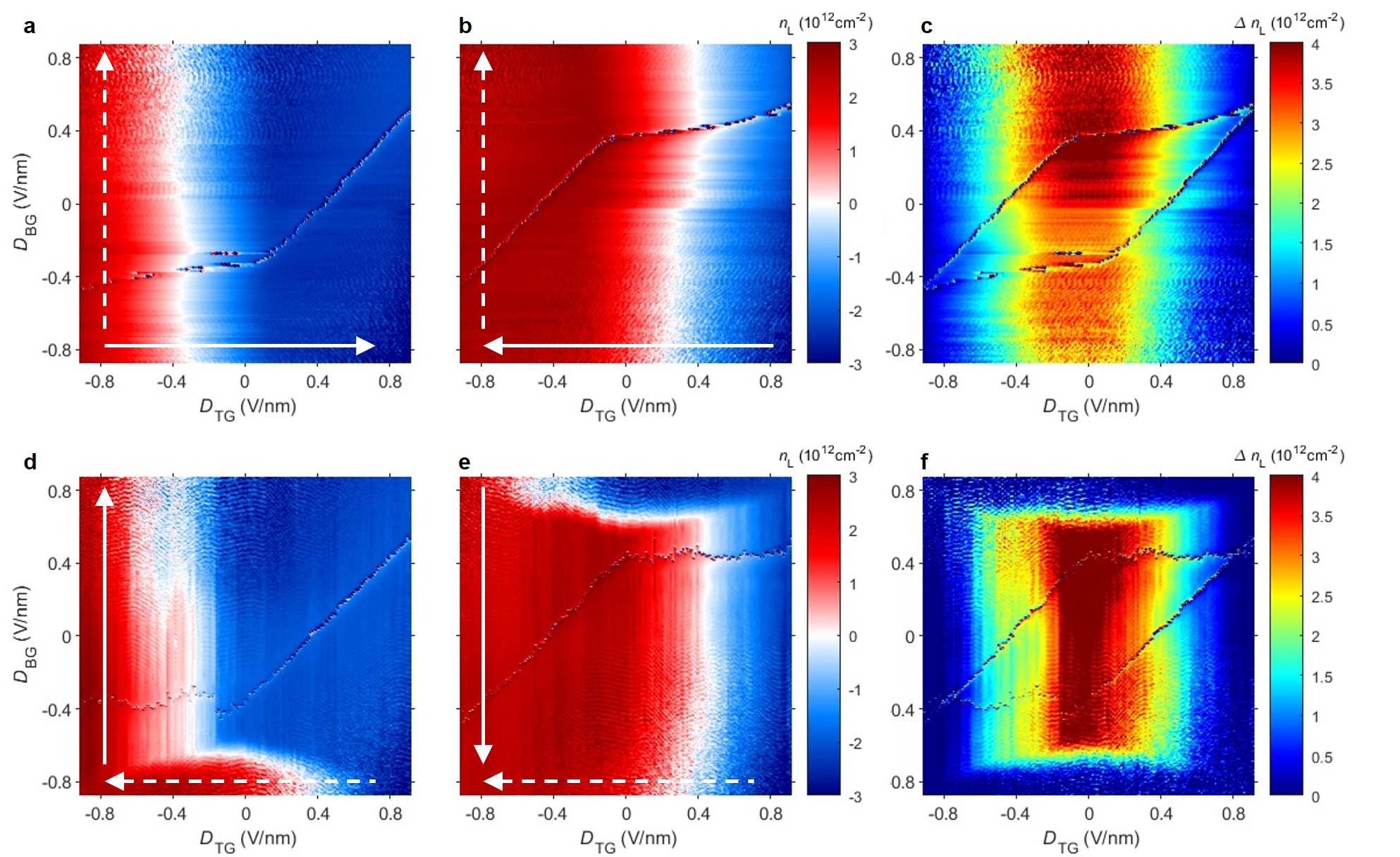}
    \caption{\textbf{$\bm n_{\text{L}}$ map over dual-gate sweep in device D1.} \textbf{a, b} Dual-gate maps of $n_{\text{L}}$ by sweeping $D_{\text{TG}}$ in the forward (\textbf{a}) and backward (\textbf{b}) directions as the fast axis, while gradually ramping up $D_{\text{BG}}$. \textbf{c} The $n_{\text{L}}$ difference  between (\textbf{b}) and (\textbf{a}), $\Delta n_{\text{L}}(D_{\text{TG}},D_{\text{BG}})$. \textbf{d, e} Dual-gate maps of $n_{\text{L}}$ by sweeping $D_{\text{BG}}$ in the forward (\textbf{d}) and backward (\textbf{e}) directions as the fast axis, with $D_{\text{TG}}$ slowly decreasing. \textbf{f}, The $n_{\text{L}}$ difference between (\textbf{e}) and (\textbf{d}), $\Delta n_{\text{L}} (D_{\text{TG}},D_{\text{BG}})$. Solid (dashed) arrows represent the fast (slow) sweep direction.}
     \label{fig:s4}
\end{figure}

For the sweep range dependence of the back gate, as shown in Fig.~4b, when $V_{\text{BG}}$ sweep range is between $\pm$90~V (bottom curve) and $\pm$10 V (second-from-top curve), $n_{\text{L}}$ changes immediately upon reversal of the back gate sweep direction. The change in $n_{\text{L}}$ under $D_{\text{BG}}$ doping then ceases once all localized states are occupied, yielding a hysteretic loop in the $D_{\text{BG}}$ sweep. In contrast, when the back gate sweep range is too small---such as 5~V (top curve in Fig.~4b)---the hysteresis in $n_{\text{L}}$ versus $D_{\text{BG}}$ vanishes, as the localized states remain unsaturated within this range. The hysteresis window sizes of $n_{\text{L}}$ are consistently $\sim 3.6 \times 10^{12}$\,cm$^{-2}$, similar to those observed in the top gate sweeps, indicating a common physical origin for the hysteresis in both $D_{\text{TG}}$ and $D_{\text{BG}}$ sweeps. 

\vspace{10mm}
\noindent \textbf{Localized charge density dependence on dual gate}

Figure~\ref{fig:s4} presents the complete dataset of the extracted localized charge density, $n_{\text{L}}$, corresponding to the measurements shown in Fig.~5. As illustrated in Fig.~\ref{fig:s4}c, the width of the hysteresis window in the $D_{\text{TG}}$ sweep remains nearly constant, regardless of variations in the back-gate displacement field, $D_{\text{BG}}$. In contrast, the hysteresis window of $n_{\text{L}}$ during $D_{\text{BG}}$ sweeps gradually narrows as $|D_{\text{TG}}|$ increases, eventually vanishing at $|D_{\text{TG}}| \approx 0.6$~V/nm. 

It is important to note that relying solely on the shift of the charge neutrality point (CNP or Dirac point)---indicated by the black lines in Fig.~\ref{fig:s4} and the $R_{xx}$ map used in Fig.~2---leads to a significant loss of information regarding the hysteretic behavior. Specifically, the Dirac point in $R_{xx}$ may shift beyond the accessible gate voltage range, and its position does not always fully capture the evolution of the hysteresis. For instance, in Figs.~\ref{fig:s4}a--c, as $|D_{\text{BG}}|$ exceeds approximately $0.4$~V/nm, the hysteresis manifested by the Dirac point shift appears to diminish and eventually vanish because the CNP moves out of the measurement window. However, the hysteresis amplitude of $n_{\text{L}}$ remains clearly discernible and nearly invariant across the same range of $D_{\text{BG}}$. This underscores the advantage of using $n_{\text{L}}$ to characterize the full extent of charge localization.

\vspace{10mm}
\large
\noindent \textbf{Supplementary Note 2: Additional data in the device R1}

\vspace{5mm}
\normalsize
\noindent \textbf{Dual-gate sweep in device R1}
\begin{figure}[ht]
    \centering
    \includegraphics[scale=1]{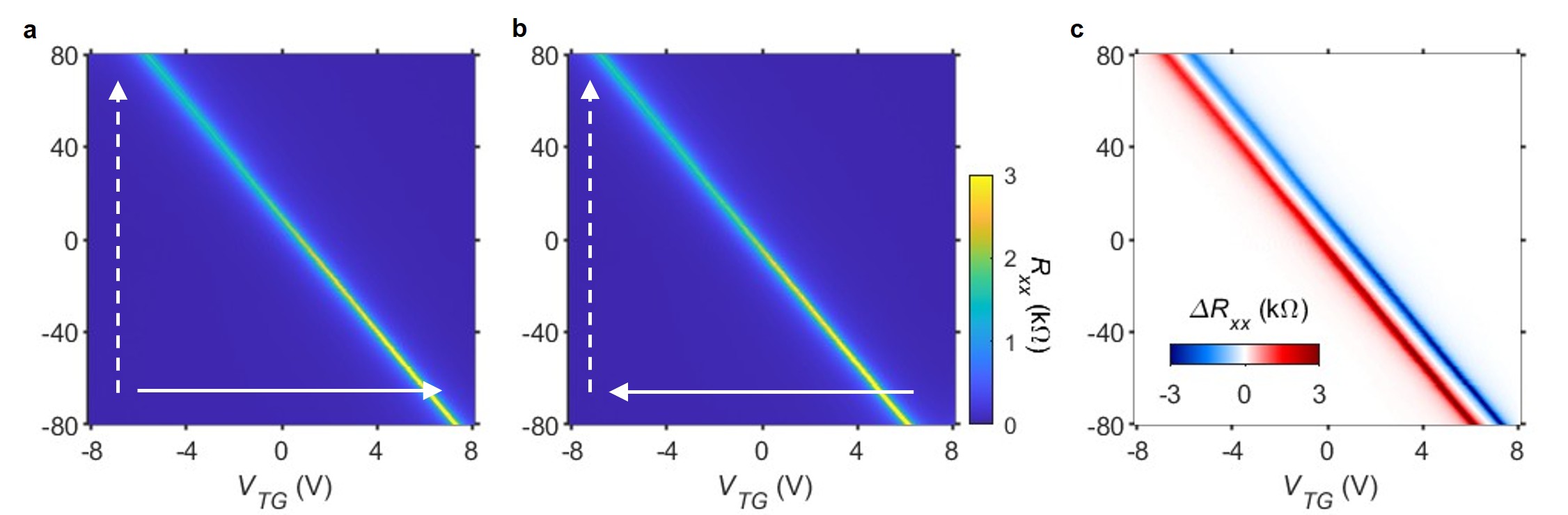}
    \caption{\textbf{Dual-gate map in device R1.} \textbf{a, b} Dual-gate maps of $R_{xx}$ by sweeping $V_{\text{TG}}$ in the forward (\textbf{a}) and backward (\textbf{b}) directions as the fast axis, while gradually ramping up $V_{\text{BG}}$. \textbf{c} The resistance difference between (\textbf{b}) and (\textbf{a}), $\Delta R_{xx}(V_{\text{TG}},V_{\text{BG}})$. Solid (dashed) arrows represent the fast (slow) sweep direction.}
     \label{fig:s5}
\end{figure}

\begin{figure}[ht]
    \centering
    \includegraphics[scale=1]{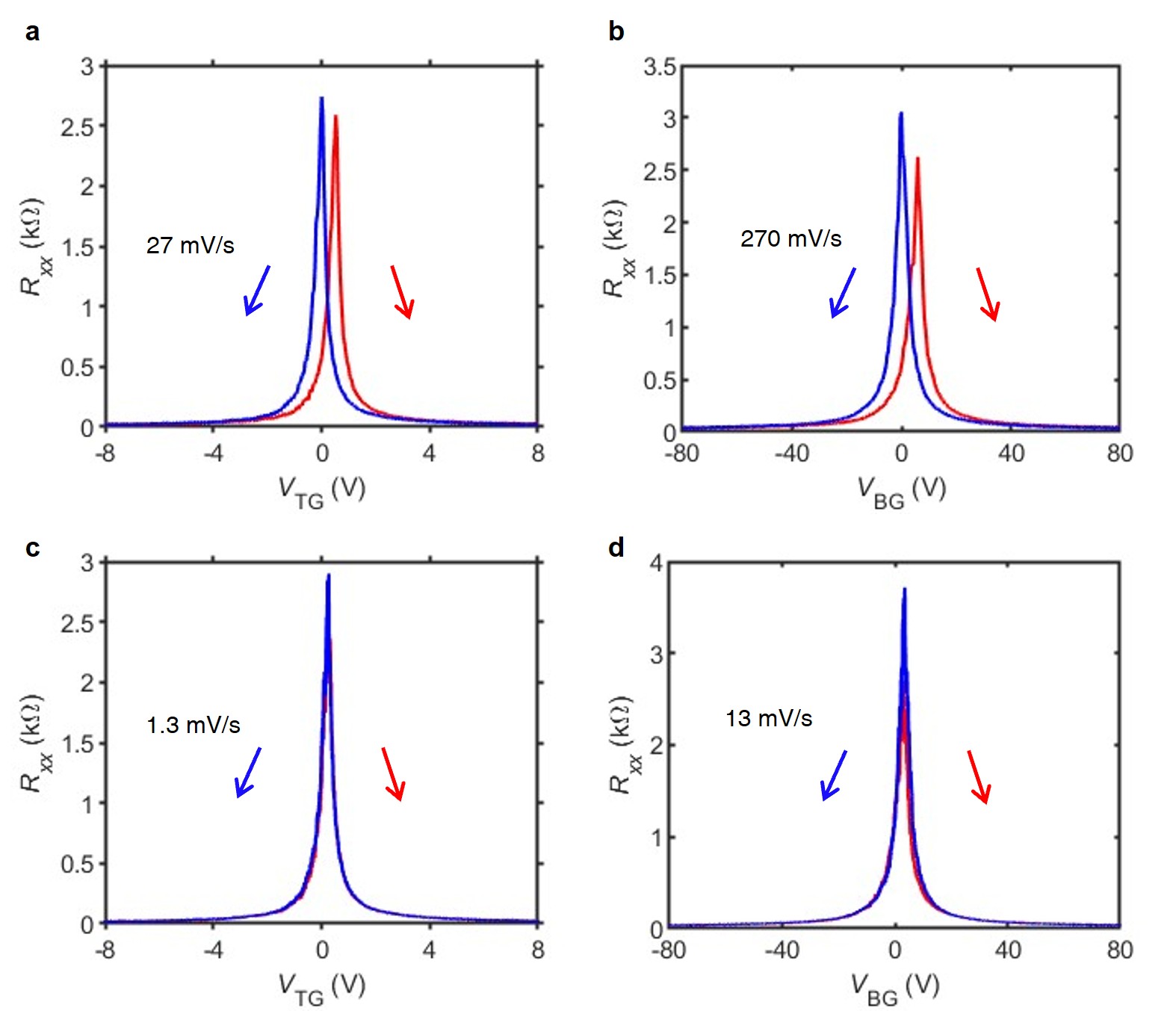}
    \caption{\textbf{Sweeping rate dependence in device R1.} \textbf{a, c} $R_{xx}$ by sweeping $V_{\text{TG}}$ in the forward (red) and backward (blue) directions with the sweeping rate of 27~mV/s (\textbf{a}) and 1.3~mV/s~(\textbf{c}). $V_{\text{BG}}$ is fixed to 0~V. \textbf{b, d} $R_{xx}$ by sweeping $V_{\text{BG}}$ in the forward (red) and backward (blue) directions with the sweeping rate of 270~mV/s (\textbf{b}) and 13~mV/s~(\textbf{d}). $V_{\text{TG}}$ is fixed to 0~V.}
     \label{fig:s6}
\end{figure}

\begin{figure}[ht]
    \centering
    \includegraphics[scale=1]{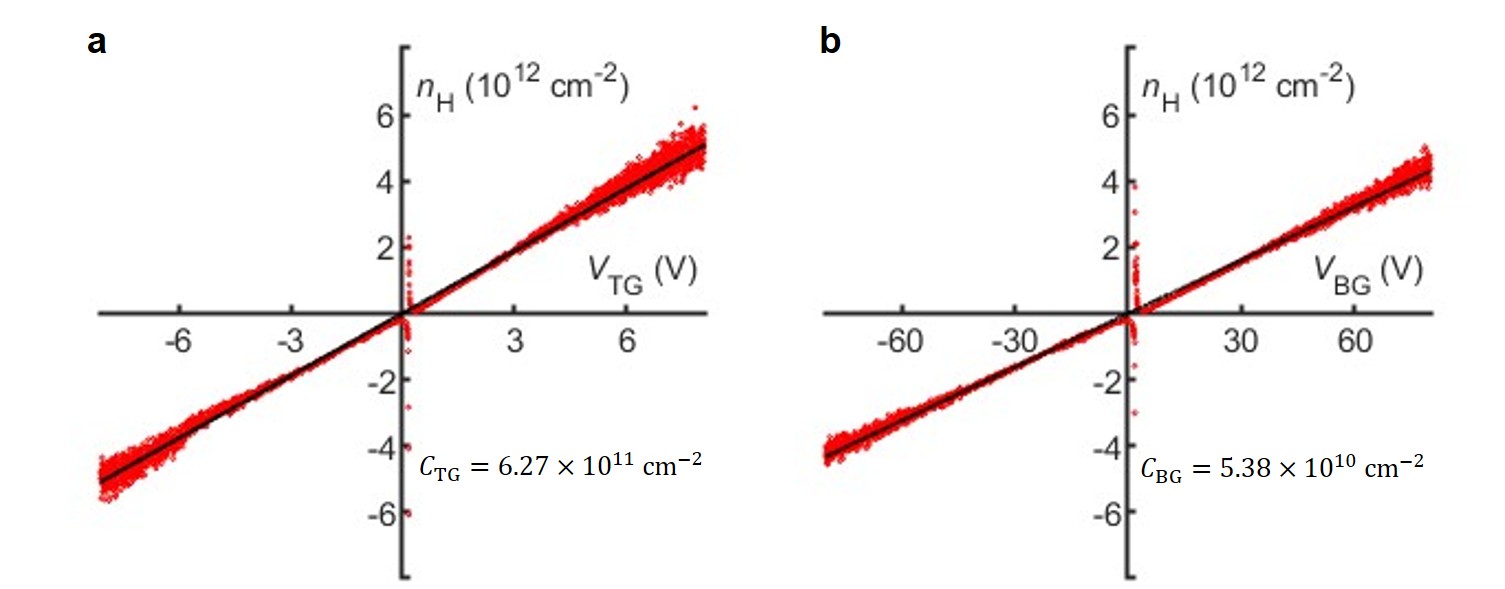}
    \caption{\textbf{Classical Hall effect in device R1.} \textbf{a} Hall carrier density $n_{\text{H}}$ as a function of top-gate voltage $V_{\text{TG}}$ (red circle). The solid black line represents a linear fit to the data, yielding $C_{\text{TG}}=6.27 \times 10^{11} \, \text{cm}^{-2}\text{V}^{-1}$. \textbf{b} Hall density $n_{\text{H}}$ plotted against back-gate voltage $V_{\text{BG}}$ (red circle). The solid black line is a linear fit yielding $C_{\text{BG}}=5.38 \times 10^{10} \, \text{cm}^{-2} \, \text{V}^{-1}$.}
     \label{fig:s7}
\end{figure}

As shown in Figs.~\ref{fig:s5}a and b, the Dirac point (resistance maximum) of the reference device R1 exhibits a straight diagonal line which is normal for dual-gate graphene devices. Also, we note that there is a small hysteresis between sweeping $V_{\text{TG}}$ in the backward and forward direction (Fig.~\ref{fig:s5}c). We attribute this small hysteresis to an extrinsic origin: surface dipole formed during the transfer process~\cite{wang2010hysteresis}, since it can be characterized by the following two properties: 1. the Dirac point is shifted to the left between forward to backward sweep direction. 2. the small hysteresis gradually decreases and even disappears after the sweeping rate is reduced. This is more clearly evidenced by the single-gate hysteresis in both top gate and back gate sweep (Fig.~\ref{fig:s6}). Since the transport properties in device R1 exhibit normal graphene properties, we use it as a reference sample for the device D1 as they share nearly same device structures. By measuring the classical Hall effect, we can extract the Hall density and the gate capacitance by $n_{\text{H}}=C_{\text{TG}} \cdot V_{\text{TG}}-n_i$, where $n_i$ is residue impurity density. As shown in Fig.~\ref{fig:s7}, top-gate capacitance is $C_{\text{TG}}=6.27 \times 10^{11} \, \text{cm}^{-2} \, \text{V}^{-1}$; and back-gate capacitance is: $C_{\text{BG}}=5.38 \times 10^{10} \, \text{cm}^{-2} \, \text{V}^{-1}$.  

\vspace{10mm}
\large
\noindent \textbf{Supplementary Note 3: Additional data in the device D2}
\vspace{5mm}
\normalsize

\begin{figure}[ht]
    \centering
    \includegraphics[scale=1]{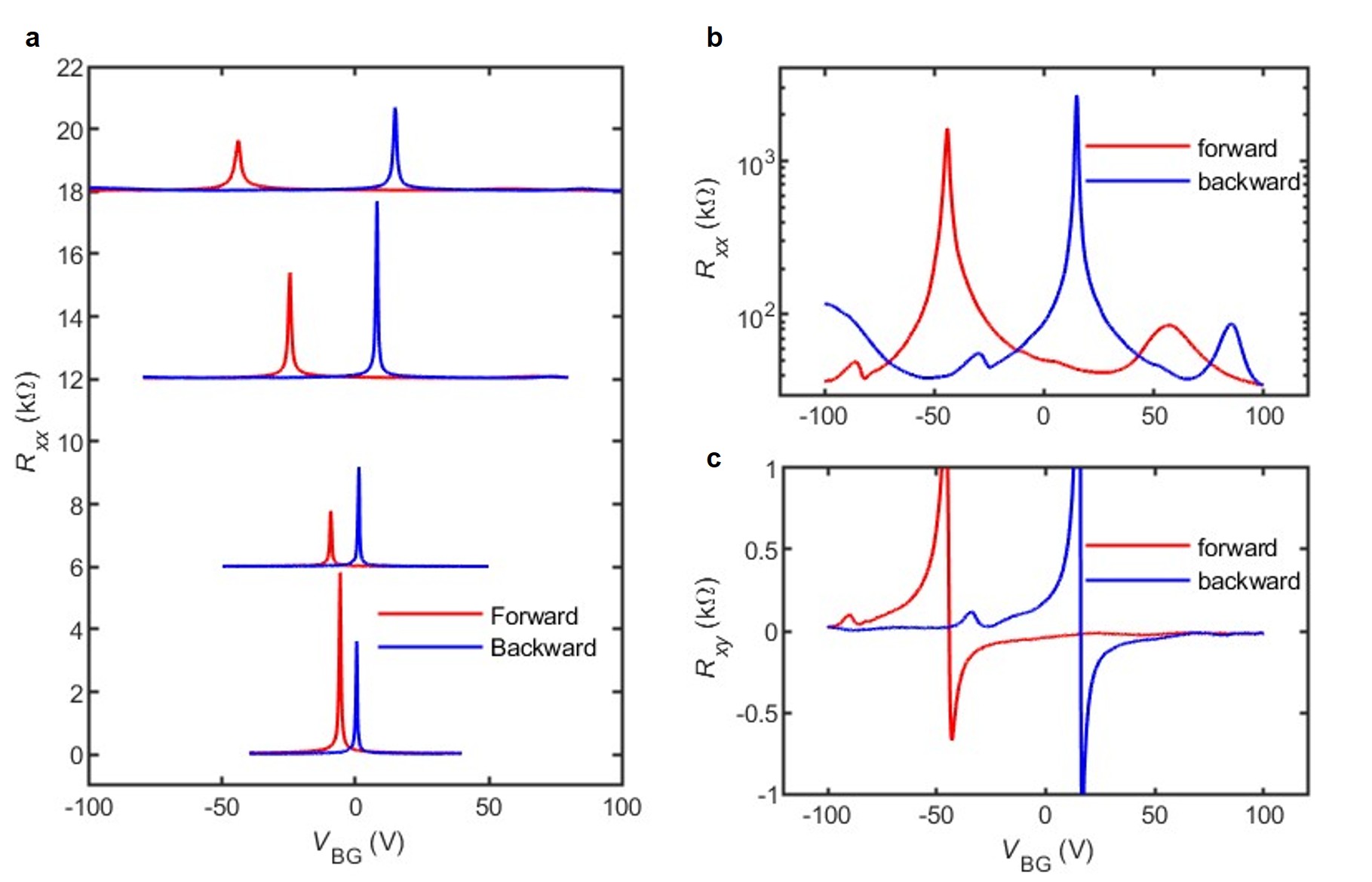}
    \caption{\textbf{Gate hysteresis in device D2.} \textbf{a} Longitudinal resistance $R_{xx}$ as a function of back-gate voltage $V_{\text{BG}}$ for forward (red) and backward (blue) sweeps. Results are shown for various sweep ranges ($\pm$40~V, $\pm$50~V, $\pm$80~V, and $\pm$100~V), with curves vertically offset by 6~k$\Omega$ for clarity. \textbf{b} $R_{xx}$ versus $V_{\text{BG}}$ for the $\pm$100~V sweep range plotted on a logarithmic scale. \textbf{c}~Transverse (Hall) resistance $R_{xy}$ versus $V_{\text{BG}}$ for the same $\pm$100~V sweep range, measured at B~=~0.5~T. All data are collected at T~=~43~K.}
     \label{fig:s8}
\end{figure}

In device D2, the trilayer hBN beneath the Hall bar contains a visible crack (Fig.~1d). A reference device, R2, was fabricated simultaneously and shares the same top and bottom hBN interfaces with the graphene; notably, R2 exhibits negligible hysteresis in both $V_{\text{TG}}$ and $V_{\text{BG}}$ sweeps.

The back-gate hysteresis in device D2 displays a sweep-range dependence similar to that observed in device D1 (see Fig.~\ref{fig:s3}b and Fig.~\ref{fig:s8}a). As the sweep range of back-gate voltage increases from $\pm$40~V to $\pm$100~V, the shift of the Dirac point between the backward and forward scans increases monotonically. Furthermore, we observed several additional $R_{xx}$ peaks during the gate sweep (see Fig.~\ref{fig:s8}b), suggesting a modification of the graphene band structure. However, these features do not resemble the typical satellite peaks resulting from an hBN/graphene moir\'{e} potential~\cite{Dean2013Hofstadter,Hunt2013massive,Yankowitz2012emergence}.

First, the peaks lack electron-hole symmetry, a behavior likely stemming from the change in gate capacitance~\cite{lin2025room_s} associated with unconventional ferroelectricity. Second, the additional peaks on the electron side are more pronounced than those on the hole side, which deviates from the general characteristics of hBN-graphene moir\'{e} potentials~\cite{Dean2013Hofstadter,Hunt2013massive}. Third, we observe peaks distinct from first-order moir\'{e} peaks, such as the feature at $\sim$100~V in the backward sweep (Fig.~\ref{fig:s8}b); however, the characteristics of this peak distinguish it from higher-order satellite moir\'{e} peaks, such as the peak amplitude and voltage distance between peaks at different orders~\cite{Lu2020high-order}. Fourth, some of the additional peaks shown in Fig.~\ref{fig:s8}b do not exhibit corresponding zero-crossings in the $R_{xy}$ measurements (see Fig.~\ref{fig:s8}c). Collectively, these discrepancies indicate that these additional peaks are specific consequences of the unconventional ferroelectricity although it can be related to moir\'{e} effects. 

Based on optical imaging of devices D2 and R2, the relative twist angle between the top hBN and the bottom trilayer hBN is estimated to be approximately $15^\circ$. The relative angle between the bottom trilayer hBN and the graphene is approximately $2^\circ$, while the angle between the bottom trilayer hBN and the thick bottom hBN is $\sim 0^\circ$. Although the proximal bottom trilayer hBN and graphene appear aligned in optical images, and transport data suggest potential moir\'{e} signatures (see Fig.~\ref{fig:s8}b, c), the substantial difference in hysteresis magnitude between D2 and R2 implies that graphene-hBN alignment alone is insufficient to generate the unusual ferroelectric effects. Furthermore, the distinct behaviors observed in device D2 compared to D1 highlight the complexity of the interfacial interactions between graphene and the specific defects within the hBN.

\vspace{10mm}
\large
\noindent \textbf{Supplementary Note 4: Additional data in the device D3}
\vspace{5mm}
\normalsize

\begin{figure}[ht]
    \centering
    \includegraphics[scale=1]{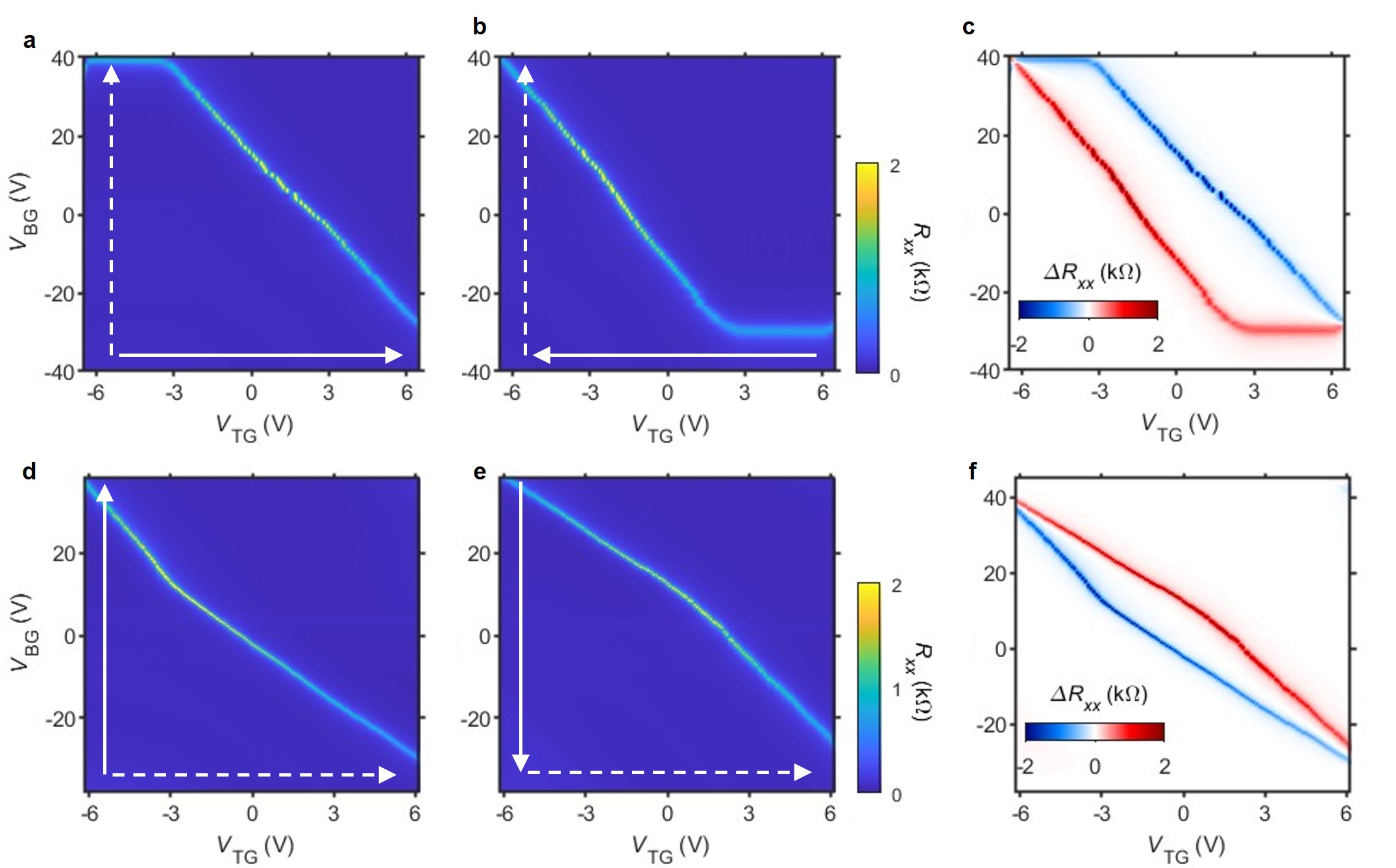}
    \caption{\textbf{Dual-gate hysteresis in device D3.} \textbf{a, b} Dual-gate maps of $R_{xx}$ obtained by sweeping $V_{\text{TG}}$ in the forward (\textbf{a}) and backward (\textbf{b}) directions (fast axis) while stepping $V_{\text{BG}}$ (slow axis). \textbf{c} The resistance difference $\Delta R_{xx}$ between (\textbf{b}) and (\textbf{a}). \textbf{d, e} Dual-gate maps of $R_{xx}$ obtained by sweeping $V_{\text{BG}}$ in the forward (\textbf{d}) and backward (\textbf{e}) directions (fast axis) while stepping $V_{\text{TG}}$ (slow axis). \textbf{f} The resistance difference $\Delta R_{xx}$ between (\textbf{e}) and (\textbf{d}). Solid (dashed) arrows indicate the fast (slow) sweep direction. Measurements were performed at $T = 5.0$~K.}
     \label{fig:s9}
\end{figure}

Device D3 features a partial trilayer hBN layer positioned between two monolayer graphene sheets beneath the Hall bar, such that the hBN does not span the entire active region (Fig.~1g). A reference device, R3, was fabricated simultaneously sharing the same top and bottom hBN interfaces; however, in R3 the inner trilayer hBN spans the entire region under the Hall bar. Notably, R3 presents negligible hysteresis in both $V_{\text{TG}}$ and $V_{\text{BG}}$ sweeps.

Based on the final optical image, all graphene/hBN interfaces are estimated to be misaligned by more than $5^\circ$, while the top and bottom graphene layers possess a relative twist angle of $1.7^\circ$, achieved via the ``cut-and-twist'' method using the tungsten tip~\cite{rao2024scratching}. As shown in Fig.~\ref{fig:s9}a--c, the top-gate voltage fails to induce charge carriers completely at different back-gate voltages when swept in the forward and backward directions, as evidenced by the horizontal Dirac point lines (resistance peaks). Differently, when the back-gate is used as the fast axis (see Fig.~\ref{fig:s9}d--f), we observe only a decrease in the slope of the Dirac point lines between the forward and backward sweeps at different top-gate voltage. 

The contrasting hysteretic behaviors observed in device D3 compared to D1 highlight the complexity of the interactions between graphene and line defects in hBN; specifically, the data suggest that the hBN defects in device D3 interact asymmetrically with the top and bottom graphene layers.

\vspace{10mm}
\large
\noindent \textbf{Supplementary Note 5: Summary of the devices with twisted hBN interface}
\normalsize

\begin{longtblr}[
  caption = {List of devices with twisted hBN interface.},
  label = {tab:S1},
]{
  colspec = {|Q[c,m,1.5cm]|Q[c,m,4cm]|Q[c,m,3cm]|Q[c,m,4cm]|},
  hlines, 
  vlines, 
  rowhead = 1, 
}
Device & hBN defects type & BG dielectric & Transport properties \\

A1 & AA$'$ (2+2, 0.5$^\circ$) & \ce{SiO2}+hBN & normal \\
A2 & AA (3+3) & \ce{SiO2}+hBN & normal \\
A3 & AA$'$ (2+2, 1$^\circ$) & \ce{SiO2}+hBN & normal \\
A4 & AA (1+1) & \ce{SiO2}+hBN & normal \\
A5 & AA (3+3, 1.6$^\circ$) & \ce{SiO2}+hBN & normal \\
A6 & AA (1+1, 1.5$^\circ$) & \ce{SiO2}+hBN & normal \\
A7 & AA$'$ (2+2, 1.5$^\circ$) & \ce{SiO2}+hBN & normal \\
A8 & AA (2+2, 58.7$^\circ$) & \ce{SiO2}+hBN & normal \\
B1 & AA (2+2, 60$^\circ$) & hBN & Sliding FE \\
B2 & AA (1+1, 0.7$^\circ$) & hBN & Sliding FE \\
B3 & AA (2+2, 59.3$^\circ$) & hBN & Satellite peaks \\
B4 & AA (1+1, 0.7$^\circ$) & hBN & Satellite peaks \\
B5 & AA (3+3, 0.7$^\circ$) & hBN & Satellite peaks \\
\end{longtblr}

\begin{figure}[ht]
    \centering
    \includegraphics[scale=1]{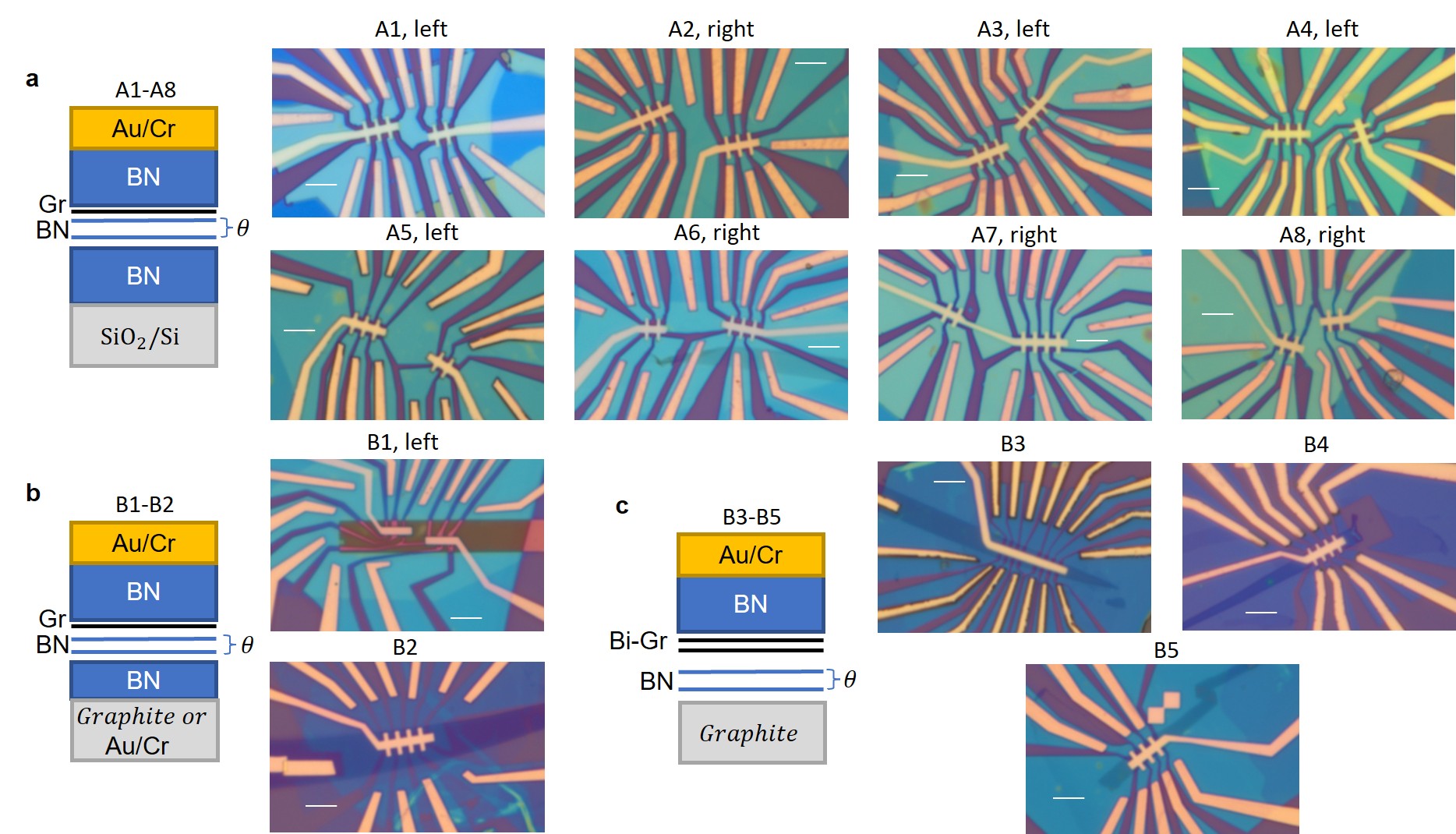}
    \caption{\textbf{Schematics and optical images of the studied devices listed in TABLE.~\Ref{tab:S1}.} Schematics and optical images of monolayer graphene devices A1--A8 with twisted hBN interface~(\textbf{a}), monolayer graphene devices B1--B2 showing sliding ferroelectricity from AA (parallel) stacked hBN~(\textbf{b}), bilayer graphene devices B3--B5 showing moir\'{e} features from AA (parallel) twisted hBN~(\textbf{c}).  For optical images with more than one device, the studied device is indicated in the caption on the top.}
     \label{fig:s10}
\end{figure}

\begin{figure}[ht]
    \centering
    \includegraphics[scale=1]{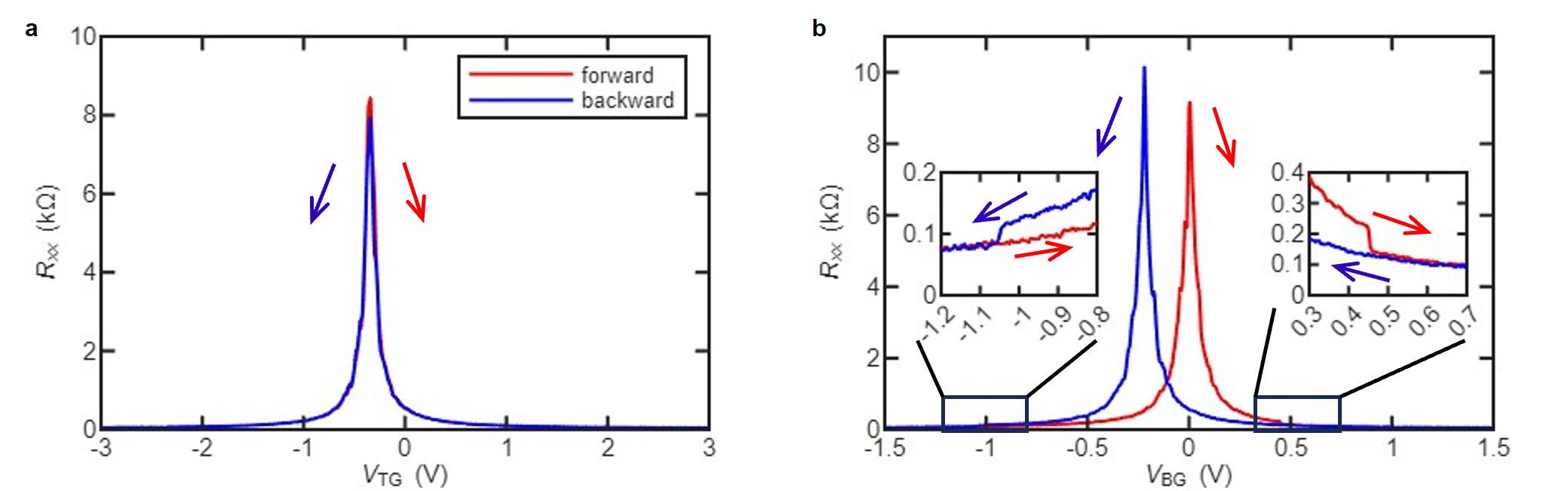}
    \caption{\textbf{Sliding ferroelectricity at a parallel-stacked hBN interface in device B2}. \textbf{a}~Longitudinal resistance $R_{xx}$ of graphene as a function of top gate voltage $V_{\text{TG}}$ in the forward (red) and backward (blue) sweep directions. \textbf{b} Longitudinal resistance $R_{xx}$ of graphene as a function of back gate voltage $V_{\text{BG}}$ in the forward (red) and backward (blue) sweep directions. The two insets in \textbf{b} show the magnified region of the marked rectangle, showing the resistance jumps at finite $V_{\text{BG}}$ due to polarization switching of the parallel-stacked hBN.}
     \label{fig:s11}
\end{figure}

\begin{figure}[ht]
    \centering
    \includegraphics[scale=0.95]{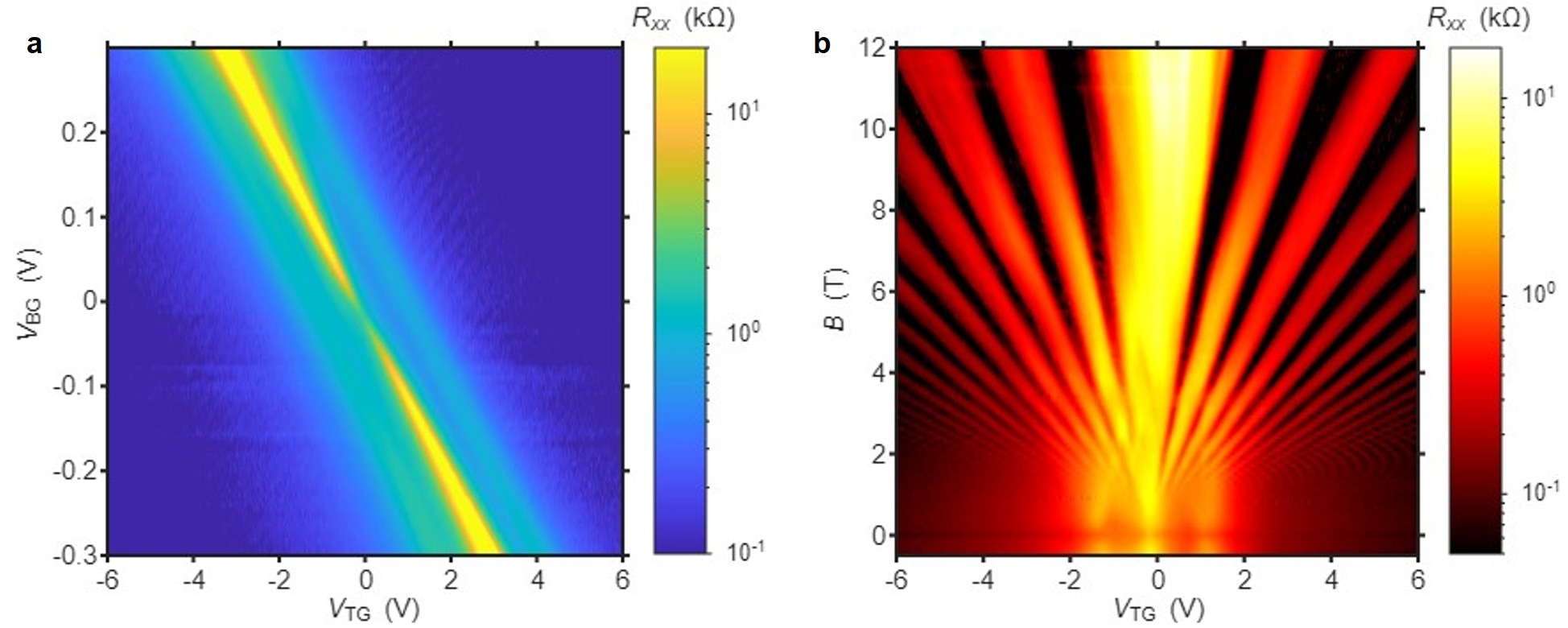}
    \caption{\textbf{Moir\'{e} potential engineering in bilayer graphene on a twisted hBN substrate (Device B5).} \textbf{a} Logarithmic scale dual-gate map of $R_{xx}$ for bilayer graphene as a function of $V_{\text{TG}}$ and $V_{\text{BG}}$. Data were acquired at a voltage bias of 100~\textmu V. \textbf{b} Logarithmic scale map of $R_{xx}$ as a function of $V_{\text{TG}}$ and magnetic field ($B$), obtained under a current bias of 10~nA.}
     \label{fig:s12}
\end{figure}

As described in the main text, we fabricated hBN-encapsulated graphene devices with twisted hBN interfaces, summarized in Table~\ref{tab:S1} and Fig.~\ref{fig:s10}. These devices include various twist angles between adjacent hBN layers (Devices A1--A8 and B1--B5). Because hBN stacking configurations can be classified as parallel (AA, AB) or anti-parallel (AA$^\prime$, AB1$^\prime$, AB2$^\prime$)~\cite{gilbert2019alternative}, small-angle twisted hBN interfaces were fabricated using a cut-and-twist method after determining the layer number of the few-layer hBN~\cite{zhang2024accurate_s}. For simplicity, we call parallel interface as AA, while anti-parallel interface as AA$^{\prime}$ in the Table.~\ref{tab:S1} and S2.

As summarized in Table~\ref{tab:S1}, graphene devices A1--A8 exhibit only conventional transport behavior, characterized by a single resistance peak and the absence of hysteresis under gate-voltage sweeps. In contrast, devices B1--B5 show either weak resistance hysteresis associated with sliding ferroelectricity (see Fig.~\ref{fig:s11}) or satellite resistance peaks indicative of moir\'{e} potential modulation (see Fig.~\ref{fig:s12}), depending on the twist angle of the parallel-stacked (AA-stacked) hBN interface. These observations are consistent with prior studies~\cite{yasuda2021stacking_s,wang2025moire_s}. In particular, the electric field on graphene created by the polarization arising from parallel-stacked hBN interfaces—whether uniform or within periodic moir\'{e} domains—decays rapidly with increasing back-gate dielectric thickness.  Consequently, a zero-angle or small-angle twisted parallel-stacked hBN interface combined with a thick \ce{SiO2} back-gate dielectric fails to induce discernible changes in the transport properties of graphene. 

Furthermore, the hysteretic samples (D1--D3) also utilize a back-gate dielectric composed of hBN and \ce{SiO2}. This implies that the origin of the observed hysteresis cannot be attributed to local polarization within the hBN or \ce{SiO2} layers independent of the graphene. Rather, there must be an interaction between the carbon atoms in the graphene and hBN defects in close proximity. We therefore conclude that hBN defects must be situated very close to the graphene layer to induce a sufficiently strong interaction.

\vspace{10mm}
\large
\noindent \textbf{Supplementary Note 6: Summary of the devices with different types of hBN defects}
\normalsize

\begin{longtblr}[
  caption = {List of devices with different types of hBN defects.},
  label = {tab:S2},
]{
  colspec = {|Q[c,m,1.5cm]|Q[c,m,4cm]|Q[c,m,3cm]|Q[c,m,4cm]|},
  hlines, 
  vlines, 
  rowhead = 1, 
}
Device & hBN defects type & BG dielectric & Transport properties \\

C1 & boundary & \ce{SiO2}+hBN & normal \\
C2 & edge & \ce{SiO2}+hBN & normal \\
C3 & {AA (3+3) edge \\ AA$'$ (3+4) edge} & \ce{SiO2}+hBN & normal \\
C4 & {AA$'$ (1+1, 60$^\circ$) edge \\ AA (1+2, 60$^\circ$) edge} & \ce{SiO2}+hBN & normal \\
C5 & {AA (3+3) edge \\ AA (2+3) edge} & \ce{SiO2}+hBN & normal \\
C6 & AA (2+1) edge & \ce{SiO2}+hBN & normal \\
C7 & AA (1+3) edge & \ce{SiO2}+hBN & normal \\
C8 & AA (3+3) edge & \ce{SiO2}+hBN & normal \\
D1 & AA boundary & \ce{SiO2}+hBN & Unconventional FE \\
D2 & crack & \ce{SiO2}+hBN & Unconventional FE \\
D3 & edge & \ce{SiO2}+hBN & Unconventional FE \\
\end{longtblr}

\begin{figure}[ht]
    \centering
    \includegraphics[scale=1]{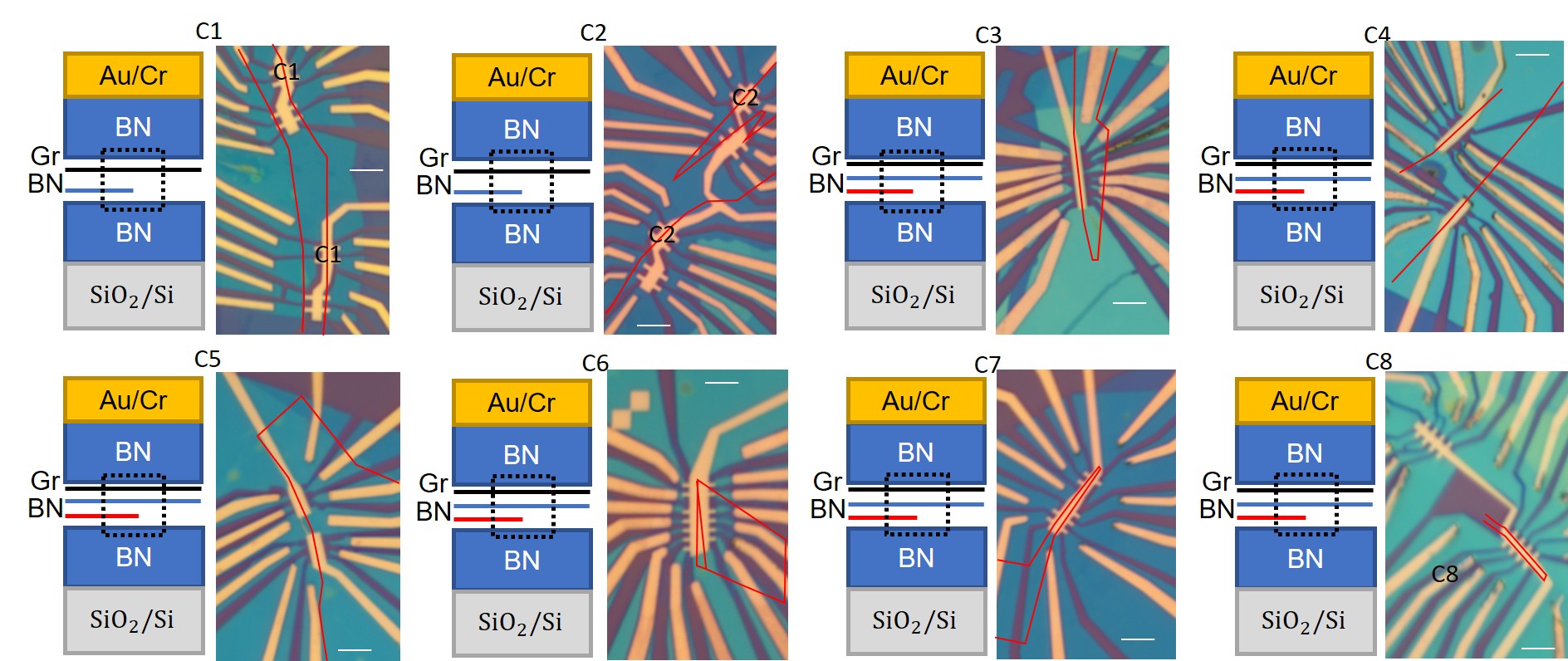}
    \caption{\textbf{Schematics and optical images of the studied devices listed in TABLE.~\Ref{tab:S2}.} Schematics and optical images of device C1--C8 with designed line defects, whose boundaries are marked in the optical images.}
     \label{fig:s13}
\end{figure}

In devices D1--D8, we attempted to produce various hBN line defects using the van der Waals (vdW) stacking method. For devices featuring natural few-layer hBN edges or steps (devices C1 and C2), both gate sweeps exhibited negligible hysteresis. As the selected flakes contained irregular (non-straight) edges or steps, they likely comprised a mixture of zigzag and armchair boundaries. This suggests that ordinary hBN boundaries are not the primary source of the unconventional ferroelectricity observed. While specific hBN boundaries have been reported to exhibit conducting properties~\cite{das2021manipulating,park2020one-dimension}—potentially acting as floating gates for charge storage~\cite{liu2021ultrafast,yu20252D,zhang2023van}—the deterministic isolation of these specific boundary types remains a significant experimental challenge.

Finally, for devices C3--C8, we investigated a configuration involving an hBN--hBN interface with a buried edge under parallel or anti-parallel stacking. Recent studies~\cite{fan2025edge_s} suggest that small-angle twisted parallel-stacked hBN can host in-plane edge polarization at the domain wall resulting from a combination of sliding ferroelectricity and piezoelectricity. The highly distorted crystal structure at these edges holds the potential to trap free charges, analogous to charged domain walls in three-dimensional ferroelectrics~\cite{sluka2013free_s}. However, graphene devices fabricated directly onto these hBN--hBN interfaces (in both parallel and anti-parallel configurations) exhibited only standard transport properties. This null result may stem from the difficulty of precisely aligning the device channel over the domain wall during fabrication; thus, further investigation is required to rigorously characterize the polarized edges at hBN interfaces before making the devices.

\newpage


\makeatletter
\let\oldendthebibliography\endthebibliography 

\renewcommand{\endthebibliography}{%
  \let\label\@gobble 
  
  \oldendthebibliography 
}
\makeatother


\newpage

\begin{thebibliography}{10}
\expandafter\ifx\csname url\endcsname\relax
  \def\url#1{\texttt{#1}}\fi
\expandafter\ifx\csname urlprefix\endcsname\relax\def\urlprefix{URL }\fi
\providecommand{\bibinfo}[2]{#2}
\providecommand{\eprint}[2][]{\url{#2}}

\bibitem{zheng2021unconventional}
\bibinfo{author}{Zheng, Z.} \emph{et~al.}
\newblock \bibinfo{title}{Unconventional ferroelectricity in moiré heterostructures}.
\newblock \emph{\bibinfo{journal}{Nature}} \textbf{\bibinfo{volume}{588}}, \bibinfo{pages}{71–76} (\bibinfo{year}{2021}).

\bibitem{niu2022giant}
\bibinfo{author}{Niu, R.} \emph{et~al.}
\newblock \bibinfo{title}{Giant ferroelectric polarization in a bilayer graphene heterostructure}.
\newblock \emph{\bibinfo{journal}{Nat. Commun.}} \textbf{\bibinfo{volume}{13}} (\bibinfo{year}{2022}).

\bibitem{zheng2023electronic}
\bibinfo{author}{Zheng, Z.} \emph{et~al.}
\newblock \bibinfo{title}{Electronic ratchet effect in a moire system: signatures of excitonic ferroelectricity}.
\newblock \emph{\bibinfo{journal}{arXiv preprint arXiv:2306.03922}}  (\bibinfo{year}{2023}).

\bibitem{Yan2023moire}
\bibinfo{author}{Yan, X.} \emph{et~al.}
\newblock \bibinfo{title}{Moiré synaptic transistor with room-temperature neuromorphic functionality}.
\newblock \emph{\bibinfo{journal}{Nature}} \textbf{\bibinfo{volume}{624}}, \bibinfo{pages}{551–556} (\bibinfo{year}{2023}).

\bibitem{klein2023electrical}
\bibinfo{author}{Klein, D.~R.} \emph{et~al.}
\newblock \bibinfo{title}{Electrical switching of a bistable moiré superconductor}.
\newblock \emph{\bibinfo{journal}{Nat. Nanotechnol.}} \textbf{\bibinfo{volume}{18}}, \bibinfo{pages}{331–335} (\bibinfo{year}{2023}).

\bibitem{chen2024selective}
\bibinfo{author}{Chen, M.} \emph{et~al.}
\newblock \bibinfo{title}{Selective and quasi-continuous switching of ferroelectric chern insulator devices for neuromorphic computing}.
\newblock \emph{\bibinfo{journal}{Nat. Nanotechnol.}} \textbf{\bibinfo{volume}{19}}, \bibinfo{pages}{962–969} (\bibinfo{year}{2024}).

\bibitem{wang2022ferroelectricity}
\bibinfo{author}{Wang, Y.} \emph{et~al.}
\newblock \bibinfo{title}{Ferroelectricity in hbn intercalated double-layer graphene}.
\newblock \emph{\bibinfo{journal}{Front. Phys. Beijing.}} \textbf{\bibinfo{volume}{17}}, \bibinfo{pages}{43504} (\bibinfo{year}{2022}).

\bibitem{zhang2024electronic}
\bibinfo{author}{Zhang, L.} \emph{et~al.}
\newblock \bibinfo{title}{Electronic ferroelectricity in monolayer graphene moiré superlattices}.
\newblock \emph{\bibinfo{journal}{Nat. Commun.}} \textbf{\bibinfo{volume}{15}}, \bibinfo{pages}{10905} (\bibinfo{year}{2024}).

\bibitem{chen2024anomalous}
\bibinfo{author}{Chen, L.} \emph{et~al.}
\newblock \bibinfo{title}{Anomalous gate-tunable capacitance in graphene moir\'e heterostructures}.
\newblock \emph{\bibinfo{journal}{arXiv preprint arXiv:2405.03976}}  (\bibinfo{year}{2024}).

\bibitem{niu2025ferroelectricity}
\bibinfo{author}{Niu, R.} \emph{et~al.}
\newblock \bibinfo{title}{Ferroelectricity with concomitant coulomb screening in van der waals heterostructures}.
\newblock \emph{\bibinfo{journal}{Nat. Nanotechnol.}} \textbf{\bibinfo{volume}{20}}, \bibinfo{pages}{346–352} (\bibinfo{year}{2025}).

\bibitem{maffione2025twist}
\bibinfo{author}{Maffione, G.} \emph{et~al.}
\newblock \bibinfo{title}{Twist-angle-controlled anomalous gating in bilayer graphene/bn heterostructures}.
\newblock \emph{\bibinfo{journal}{arXiv preprint arXiv:2506.05548}}  (\bibinfo{year}{2025}).

\bibitem{Waters2025anomalous}
\bibinfo{author}{Waters, D.} \emph{et~al.}
\newblock \bibinfo{title}{Anomalous hysteresis in graphite/boron nitride transistors}.
\newblock \emph{\bibinfo{journal}{Nano Lett.}} \textbf{\bibinfo{volume}{25}}, \bibinfo{pages}{8768–8774} (\bibinfo{year}{2025}).

\bibitem{jiang2025theinterplay}
\bibinfo{author}{Jiang, S.} \emph{et~al.}
\newblock \bibinfo{title}{The interplay of ferroelectricity and magneto-transport in non-magnetic moiré superlattices}.
\newblock \emph{\bibinfo{journal}{Nat. Commun.}} \textbf{\bibinfo{volume}{16}}, \bibinfo{pages}{5640} (\bibinfo{year}{2025}).

\bibitem{lin2025room}
\bibinfo{author}{Lin, F.} \emph{et~al.}
\newblock \bibinfo{title}{Room temperature ferroelectricity in monolayer graphene sandwiched between hexagonal boron nitride}.
\newblock \emph{\bibinfo{journal}{Nat. Commun.}} \textbf{\bibinfo{volume}{16}}, \bibinfo{pages}{1189} (\bibinfo{year}{2025}).

\bibitem{singh2025stacking}
\bibinfo{author}{Singh, A.} \emph{et~al.}
\newblock \bibinfo{title}{Stacking-induced ferroelectricity in tetralayer graphene}.
\newblock \emph{\bibinfo{journal}{arXiv preprint arXiv:2504.07935}}  (\bibinfo{year}{2025}).

\bibitem{Tu2025ferroelectric}
\bibinfo{author}{Tu, B.~Q.} \emph{et~al.}
\newblock \bibinfo{title}{Ferroelectric hysteresis in singly aligned graphene-hbn moiré superlattices}.
\newblock \emph{\bibinfo{journal}{Small}} \textbf{\bibinfo{volume}{21}}, \bibinfo{pages}{e08416} (\bibinfo{year}{2025}).

\bibitem{vizner2021interfacial}
\bibinfo{author}{Vizner~Stern, M.} \emph{et~al.}
\newblock \bibinfo{title}{Interfacial ferroelectricity by van der waals sliding}.
\newblock \emph{\bibinfo{journal}{Science}} \textbf{\bibinfo{volume}{372}}, \bibinfo{pages}{1462–1466} (\bibinfo{year}{2021}).

\bibitem{wang2022interfacial}
\bibinfo{author}{Wang, X.} \emph{et~al.}
\newblock \bibinfo{title}{Interfacial ferroelectricity in rhombohedral-stacked bilayer transition metal dichalcogenides}.
\newblock \emph{\bibinfo{journal}{Nat. Nanotechnol.}} \textbf{\bibinfo{volume}{17}}, \bibinfo{pages}{367–371} (\bibinfo{year}{2022}).

\bibitem{yasuda2021stacking}
\bibinfo{author}{Yasuda, K.}, \bibinfo{author}{Wang, X.}, \bibinfo{author}{Watanabe, K.}, \bibinfo{author}{Taniguchi, T.} \& \bibinfo{author}{Jarillo-Herrero, P.}
\newblock \bibinfo{title}{Stacking-engineered ferroelectricity in bilayer boron nitride}.
\newblock \emph{\bibinfo{journal}{Science}} \textbf{\bibinfo{volume}{372}}, \bibinfo{pages}{1458–1462} (\bibinfo{year}{2021}).

\bibitem{zhang2024accurate}
\bibinfo{author}{Zhang, T.} \emph{et~al.}
\newblock \bibinfo{title}{Accurate layer-number determination of hexagonal boron nitride using optical characterization}.
\newblock \emph{\bibinfo{journal}{Nano Lett.}} \textbf{\bibinfo{volume}{24}}, \bibinfo{pages}{14774–14780} (\bibinfo{year}{2024}).

\bibitem{Geim2013van}
\bibinfo{author}{Geim, A.~K.} \& \bibinfo{author}{Grigorieva, I.~V.}
\newblock \bibinfo{title}{Van der waals heterostructures}.
\newblock \emph{\bibinfo{journal}{Nature}} \textbf{\bibinfo{volume}{499}}, \bibinfo{pages}{419–425} (\bibinfo{year}{2013}).

\bibitem{kim2016van}
\bibinfo{author}{Kim, K.} \emph{et~al.}
\newblock \bibinfo{title}{van der waals heterostructures with high accuracy rotational alignment}.
\newblock \emph{\bibinfo{journal}{Nano Lett.}} \textbf{\bibinfo{volume}{16}}, \bibinfo{pages}{1989–1995} (\bibinfo{year}{2016}).

\bibitem{Liu2016van}
\bibinfo{author}{Liu, Y.} \emph{et~al.}
\newblock \bibinfo{title}{Van der waals heterostructures and devices}.
\newblock \emph{\bibinfo{journal}{Nat. Rev. Mater.}} \textbf{\bibinfo{volume}{1}}, \bibinfo{pages}{1–17} (\bibinfo{year}{2016}).

\bibitem{Novoselov20162D}
\bibinfo{author}{Novoselov, K.}, \bibinfo{author}{Mishchenko, A.}, \bibinfo{author}{Carvalho, A.} \& \bibinfo{author}{Castro~Neto, A.}
\newblock \bibinfo{title}{2d materials and van der waals heterostructures}.
\newblock \emph{\bibinfo{journal}{Science}} \textbf{\bibinfo{volume}{353}}, \bibinfo{pages}{aac9439} (\bibinfo{year}{2016}).

\bibitem{Wang2023towards}
\bibinfo{author}{Wang, C.}, \bibinfo{author}{You, L.}, \bibinfo{author}{Cobden, D.} \& \bibinfo{author}{Wang, J.}
\newblock \bibinfo{title}{Towards two-dimensional van der waals ferroelectrics}.
\newblock \emph{\bibinfo{journal}{Nat. Mater.}} \textbf{\bibinfo{volume}{22}}, \bibinfo{pages}{542–552} (\bibinfo{year}{2023}).

\bibitem{wang2025moire}
\bibinfo{author}{Wang, X.} \emph{et~al.}
\newblock \bibinfo{title}{Moiré band structure engineering using a twisted boron nitride substrate}.
\newblock \emph{\bibinfo{journal}{Nat. Commun.}} \textbf{\bibinfo{volume}{16}}, \bibinfo{pages}{178} (\bibinfo{year}{2025}).

\bibitem{fan2025edge}
\bibinfo{author}{Fan, W.-C.} \emph{et~al.}
\newblock \bibinfo{title}{Edge polarization topology integrated with sliding ferroelectricity in moiré system}.
\newblock \emph{\bibinfo{journal}{Nat. Commun.}} \textbf{\bibinfo{volume}{16}}, \bibinfo{pages}{3557} (\bibinfo{year}{2025}).

\bibitem{sluka2013free}
\bibinfo{author}{Sluka, T.}, \bibinfo{author}{Tagantsev, A.~K.}, \bibinfo{author}{Bednyakov, P.} \& \bibinfo{author}{Setter, N.}
\newblock \bibinfo{title}{Free-electron gas at charged domain walls in insulating \ce{BaTiO_3}}.
\newblock \emph{\bibinfo{journal}{Nat. Commun.}} \textbf{\bibinfo{volume}{4}}, \bibinfo{pages}{1808} (\bibinfo{year}{2013}).
\end{thebibliography}

\begin{thebibliography}{10}
\expandafter\ifx\csname url\endcsname\relax
  \def\url#1{\texttt{#1}}\fi
\expandafter\ifx\csname urlprefix\endcsname\relax\def\urlprefix{URL }\fi
\providecommand{\bibinfo}[2]{#2}
\providecommand{\eprint}[2][]{\url{#2}}

\bibitem{blake2007making}
\bibinfo{author}{Blake, P.} \emph{et~al.}
\newblock \bibinfo{title}{Making graphene visible}.
\newblock \emph{\bibinfo{journal}{Appl. Phys. Lett.}} \textbf{\bibinfo{volume}{91}}, \bibinfo{pages}{063124} (\bibinfo{year}{2007}).

\bibitem{jung2007simple}
\bibinfo{author}{Jung, I.} \emph{et~al.}
\newblock \bibinfo{title}{Simple approach for high-contrast optical imaging and characterization of graphene-based sheets}.
\newblock \emph{\bibinfo{journal}{Nano Lett.}} \textbf{\bibinfo{volume}{7}}, \bibinfo{pages}{3569–3575} (\bibinfo{year}{2007}).

\bibitem{ni2007graphene}
\bibinfo{author}{Ni, Z.~H.} \emph{et~al.}
\newblock \bibinfo{title}{Graphene thickness determination using reflection and contrast spectroscopy}.
\newblock \emph{\bibinfo{journal}{Nano Lett.}} \textbf{\bibinfo{volume}{7}}, \bibinfo{pages}{2758–2763} (\bibinfo{year}{2007}).

\bibitem{wang2012thickness}
\bibinfo{author}{Wang, Y.~Y.} \emph{et~al.}
\newblock \bibinfo{title}{Thickness identification of two-dimensional materials by optical imaging}.
\newblock \emph{\bibinfo{journal}{Nanotechnology}} \textbf{\bibinfo{volume}{23}}, \bibinfo{pages}{495713} (\bibinfo{year}{2012}).

\bibitem{zhang2005experimental}
\bibinfo{author}{Zhang, Y.}, \bibinfo{author}{Tan, Y.-W.}, \bibinfo{author}{Stormer, H.~L.} \& \bibinfo{author}{Kim, P.}
\newblock \bibinfo{title}{Experimental observation of the quantum hall effect and berry's phase in graphene}.
\newblock \emph{\bibinfo{journal}{Nature}} \textbf{\bibinfo{volume}{438}}, \bibinfo{pages}{201–204} (\bibinfo{year}{2005}).

\bibitem{yasuda2021stacking_s}
\bibinfo{author}{Yasuda, K.}, \bibinfo{author}{Wang, X.}, \bibinfo{author}{Watanabe, K.}, \bibinfo{author}{Taniguchi, T.} \& \bibinfo{author}{Jarillo-Herrero, P.}
\newblock \bibinfo{title}{Stacking-engineered ferroelectricity in bilayer boron nitride}.
\newblock \emph{\bibinfo{journal}{Science}} \textbf{\bibinfo{volume}{372}}, \bibinfo{pages}{1458–1462} (\bibinfo{year}{2021}).

\bibitem{wang2010hysteresis}
\bibinfo{author}{Wang, H.}, \bibinfo{author}{Wu, Y.}, \bibinfo{author}{Cong, C.}, \bibinfo{author}{Shang, J.} \& \bibinfo{author}{Yu, T.}
\newblock \bibinfo{title}{Hysteresis of electronic transport in graphene transistors}.
\newblock \emph{\bibinfo{journal}{Acs Nano}} \textbf{\bibinfo{volume}{4}}, \bibinfo{pages}{7221–7228} (\bibinfo{year}{2010}).

\bibitem{Dean2013Hofstadter}
\bibinfo{author}{Dean, C.~R.} \emph{et~al.}
\newblock \bibinfo{title}{Hofstadter’s butterfly and the fractal quantum hall effect in moiré superlattices}.
\newblock \emph{\bibinfo{journal}{Nature}} \textbf{\bibinfo{volume}{497}}, \bibinfo{pages}{598–602} (\bibinfo{year}{2013}).

\bibitem{Hunt2013massive}
\bibinfo{author}{Hunt, B.} \emph{et~al.}
\newblock \bibinfo{title}{Massive dirac fermions and hofstadter butterfly in a van der waals heterostructure}.
\newblock \emph{\bibinfo{journal}{Science}} \textbf{\bibinfo{volume}{340}}, \bibinfo{pages}{1427–1430} (\bibinfo{year}{2013}).

\bibitem{Yankowitz2012emergence}
\bibinfo{author}{Yankowitz, M.} \emph{et~al.}
\newblock \bibinfo{title}{Emergence of superlattice dirac points in graphene on hexagonal boron nitride}.
\newblock \emph{\bibinfo{journal}{Nat. Phys.}} \textbf{\bibinfo{volume}{8}}, \bibinfo{pages}{382–386} (\bibinfo{year}{2012}).

\bibitem{lin2025room_s}
\bibinfo{author}{Lin, F.} \emph{et~al.}
\newblock \bibinfo{title}{Room temperature ferroelectricity in monolayer graphene sandwiched between hexagonal boron nitride}.
\newblock \emph{\bibinfo{journal}{Nat. Commun.}} \textbf{\bibinfo{volume}{16}}, \bibinfo{pages}{1189} (\bibinfo{year}{2025}).

\bibitem{Lu2020high-order}
\bibinfo{author}{Lu, X.} \emph{et~al.}
\newblock \bibinfo{title}{High-order minibands and interband landau level reconstruction in graphene moir\'e superlattices}.
\newblock \emph{\bibinfo{journal}{Phys. Rev. B}} \textbf{\bibinfo{volume}{102}}, \bibinfo{pages}{045409} (\bibinfo{year}{2020}).
\newblock \bibinfo{note}{PRB}.

\bibitem{rao2024scratching}
\bibinfo{author}{Rao, Q.}, \bibinfo{author}{Gao, G.}, \bibinfo{author}{Wang, X.}, \bibinfo{author}{Xue, H.} \& \bibinfo{author}{Ki, D.-K. J. A.~A.}
\newblock \bibinfo{title}{Scratching lithography, manipulation, and soldering of 2d materials using microneedle probes}.
\newblock \emph{\bibinfo{journal}{Aip Adv.}} \textbf{\bibinfo{volume}{14}}, \bibinfo{pages}{015333} (\bibinfo{year}{2024}).

\bibitem{gilbert2019alternative}
\bibinfo{author}{Gilbert, S.~M.} \emph{et~al.}
\newblock \bibinfo{title}{Alternative stacking sequences in hexagonal boron nitride}.
\newblock \emph{\bibinfo{journal}{2d Mater.}} \textbf{\bibinfo{volume}{6}}, \bibinfo{pages}{021006} (\bibinfo{year}{2019}).

\bibitem{zhang2024accurate_s}
\bibinfo{author}{Zhang, T.} \emph{et~al.}
\newblock \bibinfo{title}{Accurate layer-number determination of hexagonal boron nitride using optical characterization}.
\newblock \emph{\bibinfo{journal}{Nano Lett.}} \textbf{\bibinfo{volume}{24}}, \bibinfo{pages}{14774–14780} (\bibinfo{year}{2024}).
\newblock \bibinfo{note}{Doi: 10.1021/acs.nanolett.4c04241}.

\bibitem{wang2025moire_s}
\bibinfo{author}{Wang, X.} \emph{et~al.}
\newblock \bibinfo{title}{Moiré band structure engineering using a twisted boron nitride substrate}.
\newblock \emph{\bibinfo{journal}{Nat. Commun.}} \textbf{\bibinfo{volume}{16}}, \bibinfo{pages}{178} (\bibinfo{year}{2025}).

\bibitem{das2021manipulating}
\bibinfo{author}{Das, B.} \emph{et~al.}
\newblock \bibinfo{title}{Manipulating edge current in hexagonal boron nitride via doping and friction}.
\newblock \emph{\bibinfo{journal}{Acs Nano}} \textbf{\bibinfo{volume}{15}}, \bibinfo{pages}{20203–20213} (\bibinfo{year}{2021}).

\bibitem{park2020one-dimension}
\bibinfo{author}{Park, H.~J.} \emph{et~al.}
\newblock \bibinfo{title}{One-dimensional hexagonal boron nitride conducting channel}.
\newblock \emph{\bibinfo{journal}{Sci. Adv.}} \textbf{\bibinfo{volume}{6}}, \bibinfo{pages}{eaay4958} (\bibinfo{year}{2020}).

\bibitem{liu2021ultrafast}
\bibinfo{author}{Liu, L.} \emph{et~al.}
\newblock \bibinfo{title}{Ultrafast non-volatile flash memory based on van der waals heterostructures}.
\newblock \emph{\bibinfo{journal}{Nat. Nanotechnol.}} \textbf{\bibinfo{volume}{16}}, \bibinfo{pages}{874–881} (\bibinfo{year}{2021}).

\bibitem{yu20252D}
\bibinfo{author}{Yu, X.} \emph{et~al.}
\newblock \bibinfo{title}{2d materials-based flash memory device: mechanism, structure, application}.
\newblock \emph{\bibinfo{journal}{Mater. Horiz.}} \textbf{\bibinfo{volume}{12}}, \bibinfo{pages}{8409–8429} (\bibinfo{year}{2025}).

\bibitem{zhang2023van}
\bibinfo{author}{Zhang, Q.} \emph{et~al.}
\newblock \bibinfo{title}{Van der waals materials-based floating gate memory for neuromorphic computing}.
\newblock \emph{\bibinfo{journal}{Chip}} \textbf{\bibinfo{volume}{2}}, \bibinfo{pages}{100059} (\bibinfo{year}{2023}).

\bibitem{fan2025edge_s}
\bibinfo{author}{Fan, W.-C.} \emph{et~al.}
\newblock \bibinfo{title}{Edge polarization topology integrated with sliding ferroelectricity in moiré system}.
\newblock \emph{\bibinfo{journal}{Nat. Commun.}} \textbf{\bibinfo{volume}{16}}, \bibinfo{pages}{3557} (\bibinfo{year}{2025}).

\bibitem{sluka2013free_s}
\bibinfo{author}{Sluka, T.}, \bibinfo{author}{Tagantsev, A.~K.}, \bibinfo{author}{Bednyakov, P.} \& \bibinfo{author}{Setter, N.}
\newblock \bibinfo{title}{Free-electron gas at charged domain walls in insulating \ce{BaTiO_3}}.
\newblock \emph{\bibinfo{journal}{Nat. Commun.}} \textbf{\bibinfo{volume}{4}}, \bibinfo{pages}{1808} (\bibinfo{year}{2013}).

\end{thebibliography}
\end{document}